\documentclass[prl,aps,twocolumn]{revtex4-1}

\usepackage[utf8]{inputenc}
\usepackage[T1]{fontenc}

\usepackage{placeins}
\usepackage{lipsum}

\usepackage{bm}
\usepackage{amsmath}
\usepackage{amssymb}
\usepackage{upgreek}
\usepackage{subfigure}

\usepackage{siunitx}
\usepackage[version=4]{mhchem}

\usepackage{color}
\usepackage{graphicx}
\usepackage{wrapfig}
\usepackage{epsfig}

\usepackage{makecell}

\usepackage[unicode=true,colorlinks=true,citecolor=blue]{hyperref}

\usepackage{hyphenat}
\begin{document}

\title{Terahertz Magnetospectroscopy of Cyclotron Resonances from Topological Surface States in Thick Films of \texorpdfstring{\ce{Cd_xHg_{1-x}Te}}{CdHgTe}
}

\author{%
  M.~Otteneder\textsuperscript{1},
  D.~Sacré\textsuperscript{1},
  I.~Yahniuk\textsuperscript{2},
  G.\,V.~Budkin\textsuperscript{3},
  K.~Diendorfer\textsuperscript{1},
  D.\,A.~Kozlov\textsuperscript{4},
  I.\,A.~Dmitriev\textsuperscript{1,3},\\
  N.\,N. Mikhailov\textsuperscript{4},
  S.\,A. Dvoretsky\textsuperscript{4},
  S.\,A.~Tarasenko\textsuperscript{3},
  V.\,V.~ Bel'kov\textsuperscript{3},
  W. Knap\textsuperscript{2}, 
  and
  S.\,D.~Ganichev\textsuperscript{1}
}

\affiliation{%
  \mbox{\textsuperscript{1}\,Terahertz Center, University of Regensburg, 93040 Regensburg, Germany}\\
  \mbox{\textsuperscript{2}\,CENTERA Laboratories, Institute of High Pressure Physics, Polish Academy of Sciences, PL-01142 Warsaw, Poland}\\
  \mbox{\textsuperscript{3}\,Ioffe Institute, 194021 St. Petersburg, Russia}\\
  \mbox{\textsuperscript{4}\,Rzhanov Institute of Semiconductor Physics, 630090 Novosibirsk, Russia}
}

\begin{abstract}
We present studies of the cyclotron resonance (CR) in thick \texorpdfstring{\ce{Cd_{x}Hg_{1-x}Te}}{CdHgTe} films with different cadmium concentrations corresponding to inverted and normal band order, as well as to an almost linear energy dispersion. Our results demonstrate that formation of two-dimensional topological surface states requires sharp interfaces between layers with inverted and normal band order, in which case the corresponding CR is clearly observed for the out-of-plane orientation of magnetic field, but does not show up for an in-plane orientation. By contrast, all samples having more conventional technological design with smooth interfaces (i.e., containing regions of \texorpdfstring{\ce{Cd_{x}Hg_{1-x}Te}}{CdHgTe} with gradually changing Cd content $x$) show equally pronounced CR in both in-plane and out-of-plane magnetic field revealing that CR is excited in three-dimensional states. Modeling of the surface states for different film designs supports our main observations. In all samples, we observe additional broad helicity-independent resonances which are attributed to photo-ionization and magnetic freeze-out of impurity states.
\end{abstract}

\maketitle   

\section{Introduction}
\label{introduction}

The discovery of quantum spin Hall effect in HgTe quantum wells~\cite{Bernevig2006,Koenig2008} and of topological insulators~\cite{Hasan2010,Moore2010,Qi2011} stimulated rapidly growing interest to fundamental properties and possible applications of HgTe-based materials.
An attractive feature of these materials is that the inverted band order, crucial for the formation of time-reversal symmetry protected gapless states, can be obtained in several different ways. 
In particular, the topological phase transition can be realized via variation of the thickness of HgTe quantum wells, by changing temperature, or by applying strain. This, together with a high quality of materials grown by molecular-beam-epitaxy (MBE), provides unique opportunities to study helical Dirac fermions (for reviews, see e.g.~\cite{Qi2011,Ortmann2015})
as well as induced superconductivity and phase-controlled Josephson  junctions~\cite{Maier2012,Ren2019}. 
Owing to their specific energy dispersion, HgTe quantum wells and strained bulk films have revealed fascinating effects in magneto-transport ~\cite{Koenig2008,Bruene2011} (for recent achievements see \cite{Nowack2013,Kozlov2014,Bruene2014,Olshanetsky2015,Ma2015,Kozlov2016,Inhofer2017,Maier2017,Imhof2018,Mahler2019}), as well as in 
 magneto- and terahertz-spectroscopy~\cite{Kvon2012,Zholudev2012,Olbrich2013,Shuvaev2013,Pakmehr2014,Dantscher2015,Shuvaev2016,Dantscher2017,Dziom2017,Kadykov2018,Gospodaric2019}. 

More recently, a considerable attention has been attracted to bulk \texorpdfstring{\ce{Cd_{x}Hg_{1-x}Te}}{CdHgTe} films with the Cd content $x$ equal or lower than the critical one, $x_c$, defining the phase transition from inverted to  non-inverted band order~\cite{Berchenko1976}. This interest has been driven by observation of three-dimensional massless Kane fermions in films with $x=x_c$, see Ref.~\cite{Orlita2014}, followed by detailed studies using magneto- and terahertz-spectroscopy~\cite{Teppe2016,Yavorskiy2018,But2019}.
Topological insulators based on bulk films with Cd concentration below the temperature dependent critical $x_c$ remain less studied so far~\cite{Tomaka2017,Galeeva2018}. 
 These materials, however, have a number of advantages with respect to HgTe TIs. As demonstrated in Ref.~\cite{Tomaka2017}, the topologically protected surface states in \texorpdfstring{\ce{Cd_{x}Hg_{1-x}Te}}{CdHgTe} are characterized by: (i) two times higher Dirac fermion velocity (approximately the same as in graphene) than in pure HgTe and (ii) a larger band gap and a higher position of the Dirac point on the energy scale than that obtained in strained HgTe films. These significant advantages make TIs based on HgCdTe alloys  promising for future applications.

\begin{table*}[tb]
	\centering
\begin{tabular*}{\textwidth}{@{\extracolsep{\fill}} c c c c c c c}
	\hline
	\hline
	sample & Cd content, $x$ & \makecell{band\\ structure} & top interface &  $n_\mathrm{s}\,\left(10^{11}\,\text{cm}^{-2}\right)$ & $n\,\left(10^{14}\,\text{cm}^{-3}\right)$ \\ 
	\hline
	\hline 
\#A
	& 0.151 & inverted & smooth & 2.88 & 7.73 \\

\#B
	& 0.150 & inverted & smooth & 2.21 & 6.05\\	

\#C
	& 0.151 & inverted & sharp & 3.33 & 9.65\\	

\#D
	& 0.179 & normal & smooth & 4.19 & 3.99 \\

\#E
	& 0.223 & normal & smooth & 2.44 & 3.35 \\

	\hline
	\hline
\end{tabular*}
\caption{Basic parameters of the investigated samples \#A - \#E including the Cd concentration $x$ and the corresponding type of the band order in the flat region, type of interfaces, as well as the electron densities (the average 3D electron density $n$ and the corresponding effective sheet density $n_s$) obtained from low-frequency magnetotransport measurements at $T=4.2$~K.
}
\label{transporttable}
\end{table*}

Here we present a detailed study of terahertz cyclotron resonances (CR) of thick \texorpdfstring{\ce{Cd_{x}Hg_{1-x}Te}}{CdHgTe} undoped films observed in transmission, photocurrent~\cite{Dantscher2015,Otteneder2018,Candussio2019}, and photoconductivity. 
The films under study are characterized by different value of $x$, corresponding to both normal and inverted band order.
In the samples characterized by inverted band order two CRs have been observed in the Faraday configuration. Helicity dependencies of the resonance amplitudes together with their different positions allow us to conclude that the resonances originate from two kinds of negatively charged carriers with different masses. The corresponding cyclotron masses (below $\text{0.01} m_0$), obtained from positions of the CRs, may be attributed to both bulk electrons and electrons in topological surface states, which are expected to exist in the films with inverted band order.
Our central observations come from CR measurements in Voigt configuration, with magnetic field applied parallel to the sample surface. These measurements allow us to distinguish between two-dimensional (2D) and three-dimensional (3D) behaviors of carriers responsible for the observed resonances. In this geometry, we obtain qualitatively different results for films with abruptly changing and gradually varying Cd concentration $x$ at the interfaces between layers with inverted and normal band order. While samples with smoothly varying $x$ manifest almost identical resonant behavior in Faraday and Voigt configurations, in the sample with an abrupt transition to the cap layer one of the electron CRs disappears in the Voigt geometry. This observation clearly indicates a 2D nature of one group of electrons, apparently located at the abrupt interface and associated with a topological surface state. These conclusions are supported by theoretical modelling of the edge states in films with abrupt and smooth interfaces. Therefore, the presence of a sharp interface appears to be a crucial technological requirement for studies of fundamental topological properties of such structures. 
In samples with normal band order we detect only one cyclotron resonance, with the CR mass close to that calculated for the bulk material. 
In all studied films we observe an additional broad resonance located in higher magnetic fields. This helicity-independent resonance manifests slower kinetics and decays rapidly with temperature. We assign this resonance to photo-ionization and magnetic freeze-out of impurity states.
\section{Samples}
\label{samples_methods}

\begin{figure}[tb]
	\centering
	\includegraphics[width=\linewidth]{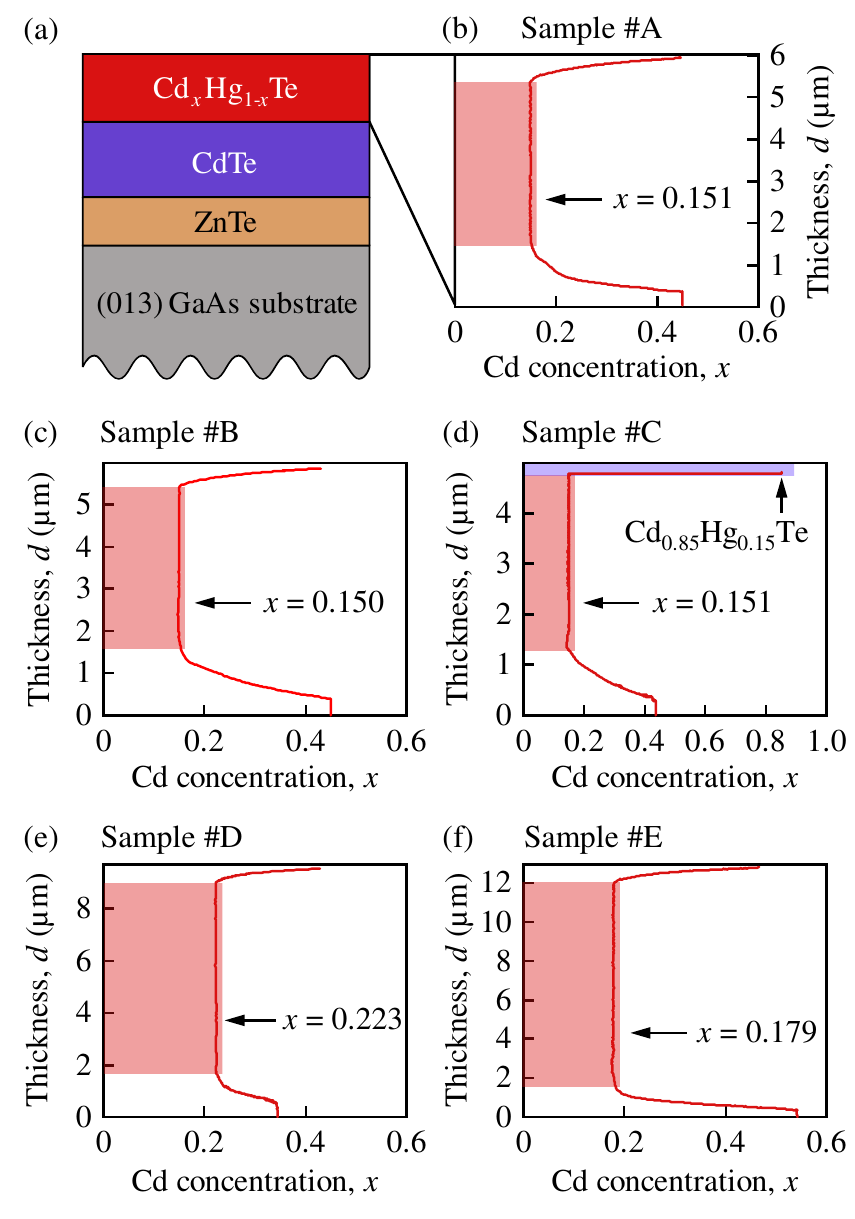}
	\caption{(a) Schematic layer structure of all investigated \texorpdfstring{\ce{Cd_{x}Hg_{1-x}Te}}{CdHgTe} films. Panels (b), (c), (d), (e), and (f) display the Cd concentration profile $x(d)$ in the top layer as a function of the distance $d$ from the top of preceding CdTe layer. The values of $x$ in the flat region between interfaces are indicated in the corresponding panels. Note that unlike other structures with smoothly varying $x(d)$, the sample \#C in panel (d) has a sharp interface between the flat region made of \texorpdfstring{\ce{Cd_{0.15}Hg_{0.85}Te}}{CdHgTe} with inverted band order and a 30 nm cap layer made of \texorpdfstring{\ce{Cd_{0.85}Hg_{0.15}Te}}{CdHgTe} with normal band order.
}
	\label{fig1}
\end{figure}

\noindent The investigated samples are  molecular beam epitaxy grown \texorpdfstring{\ce{Cd_{x}Hg_{1-x}Te}}{CdHgTe} films having a common general layer structure as sketched in figure~\ref{fig1}. For each sample, a $\SI{30}{\nano\metre}$ ZnTe buffer layer was grown on top of a (013)-oriented GaAs substrate followed by a $\SI{6}{\micro\metre}$ layer of CdTe. After that, \texorpdfstring{\ce{Cd_{x}Hg_{1-x}Te}}{CdHgTe} films with different composition (as described below) and thickness ranging in between 5 and 12~$\SI{}{\micro\metre}$ were fabricated. All wafers were cleaved into square $5\times \SI{5}{\milli\metre\squared}$ samples. Ohmic indium contacts have been soldered to the sample edges and corners.

Figures \ref{fig1} (b) - (f) show the cadmium concentration $x$ in \texorpdfstring{\ce{Cd_{x}Hg_{1-x}Te}}{CdHgTe} films selected for the presentation as a function of the distance $d$ from the top of the preceding CdTe layer. In the initial step of growth of the film ($d\lesssim\SI{1.5}{\micro\metre}$), the cadmium concentration $x$ was gradually decreased in order to reduce the strain related to a lattice mismatch, as well as to avoid possible interface disorder effects. Such smooth interfaces with varying $x$ are conventionally used to improve opto-electronic and transport properties of \texorpdfstring{\ce{Cd_{x}Hg_{1-x}Te}}{CdHgTe} films, which are widely used for detection of infrared radiation.
The region of varying $x$ is followed by a wide region with constant $x=0.15$ (samples \#A, \#B, and \#C), $x=0.18$ (sample \#D), or $x=0.22$ (sample \#E) which will be called the flat region in the following.
After that a cap layer with gradually increasing Cd content was grown in all samples except for the sample \#C. In the sample \#C, the flat region with $x=0.15$ was capped by a $\SI{30}{\nano\meter}$ layer of \texorpdfstring{\ce{Cd_{0.85}Hg_{0.15}Te}}{CdHgTe}, which resulted in a sharp boundary between the flat region with an inverted band order and the cap layer with a normal band order.

\begin{figure}[tb]
	\centering
	\includegraphics[width=\linewidth]{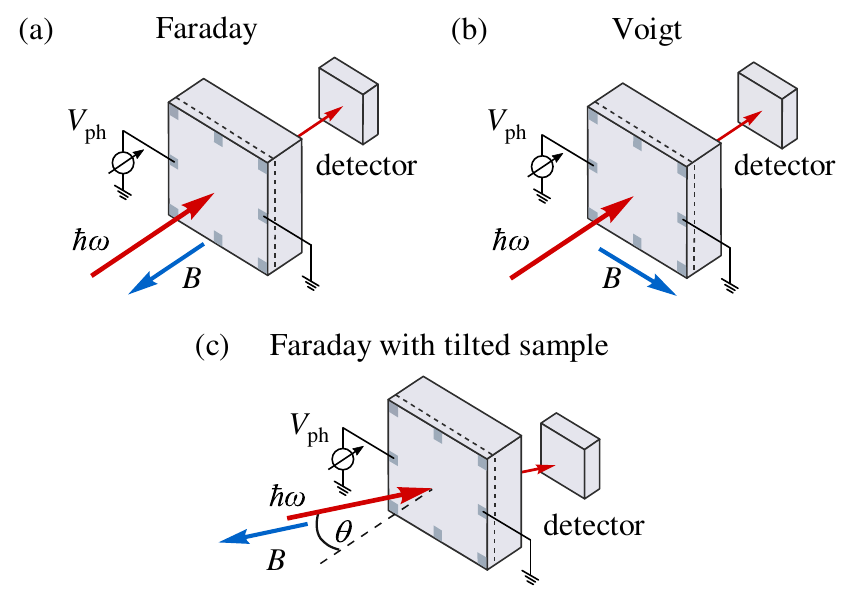}
	\caption{Experimental setups for transmission and photovoltage measurements: (a) Faraday configuration with both magnetic field and the incident THz beam oriented along the normal to the sample surface; (b) Voigt configuration with an in-plane magnetic field; (c) Faraday configuration with the sample being tilted by an angle $\theta$ with respect to the common direction of the laser beam and magnetic field.}
	\label{fig2}
\end{figure}

\begin{figure}[tb]
	\centering
	\includegraphics[width=\linewidth]{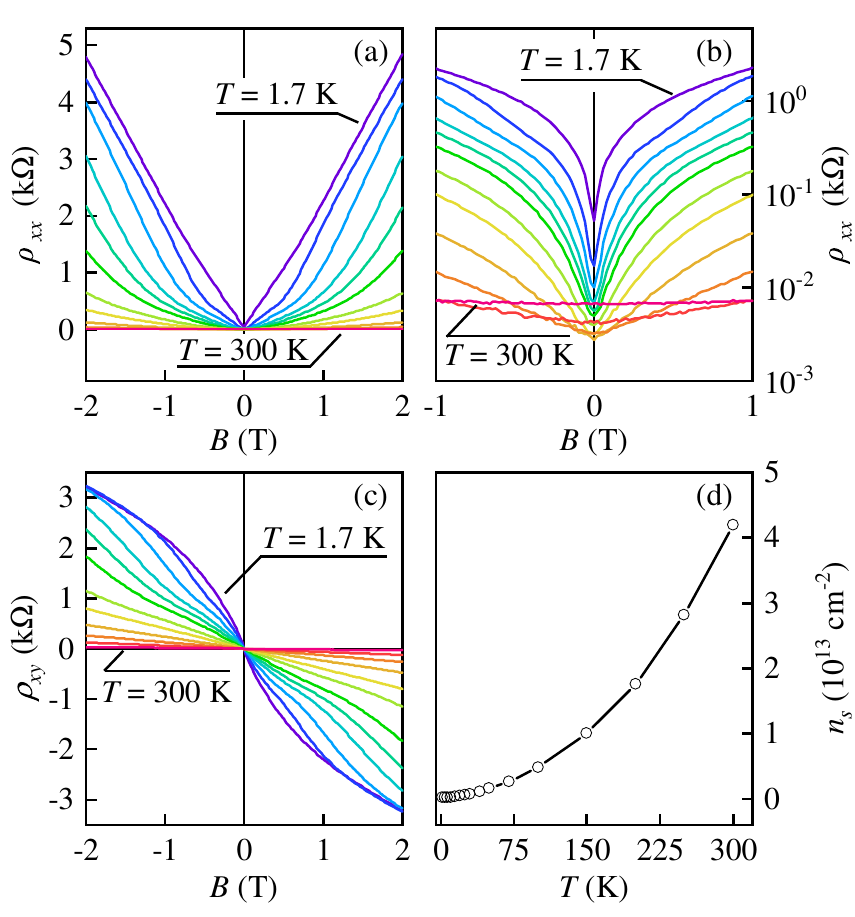}
	\caption{Magnetotransport data recorded on a Hall bar made from sample \#A with $x=0.15$ at different temperatures, $T=1.7$, 10, 15, 20, 25, 30, 40, 50, 70, 100, 150, and 300~K. Panels (a) and (b): the longitudinal sheet resistivity $\rho_{xx}$ in linear and logarithmic scale, respectively.  Panel (c): the Hall resistivity $\rho_{xy}$. Panel (d): The effective surface electron density $n_s(T)$ obtained from the linear slope of $\rho_{xy}$ at $B=0$ in panel (c).}
	\label{fig3}
\end{figure}

All samples were characterized by transport measurements performed using standard low-frequency lock-in technique in the temperature range from 2 to $\SI{300}{\kelvin}$ and an out-of-plane magnetic field up to $\SI{10}{\tesla}$. 
These studies established that all samples exhibit similar transport behavior that is not sensitive to the existence of band inversions and smooth gradients of $x$ at the interfaces. At low temperatures, all samples show strong positive magnetoresistance and a non-linear Hall effect as illustrated in figure~\ref{fig3} showing data for the sample \#A. The slope of the Hall resistivity at low magnetic field corresponds to negatively charged carriers (electrons), while at strong field the Hall slope typically changes to a hole-like. Such behavior is typical for electron-hole systems \cite{Kvon2008} in conditions when the hole density exceeds the electron density (though both have the same order of magnitude), while electrons are characterized by a significantly higher mobility. The values of density and mobility for different types of carriers can be extracted using the two-component classical Drude model, which was successfully implemented for thin HgTe films \cite{Kozlov2014,Candussio2019,Savchenko2019}. The values obtained from such modeling are shown in Table \ref{transporttable}. Note that this technique does not allow one to distinguish possible different groups of electrons (or holes) and yields the total electron and hole densities and average mobilities only. We also mention that agreement between measured transport data and fits using the two-component model was not very good in the region of low magnetic fields. Therefore, in figure~\ref{fig3}~(d) we present the results of alternative simpler treatment within the single-component Drude model, where the effective sheet density of electrons in the film is obtained from the linear slope of the Hall resistivity at low magnetic fields.

Overall, transport measurements demonstrate that at low temperatures the electron density is rather small and lies in a range of $n_s=2 \div \SI{4e11}{\per\centi\metre\squared}$. 
The hole density slightly exceeds the electron density but has the same order of magnitude. The holes do not contribute to the CRs observed, so their properties will not be discussed in detail. However, they play an important role in keeping the electro-neutrality. At higher temperatures both electron and hole densities increase mainly because of usual temperature smearing of the energy distribution, but temperature variation of the bulk energy gap may also play a role \cite{Teppe2016}.
Typically, at $\SI{77}{\kelvin}$ the electron density reaches  $n_s = 2\div \SI{3e12}{\per\centi\metre\squared}$ for all samples, while at $\SI{300}{\kelvin}$ the density rises to $n_s = \SI{3e13}{} \div \SI{1.2e14}{\per\centi\metre\squared}$. 
It is worth noting that at the liquid nitrogen temperature every sample shows an order of magnitude increase of density in comparison to low temperatures, and, at the same time, the values of $n_s$ are very similar for all samples independent of the value of the band gap in the flat region. Such behavior suggests that at $T<\SI{77}{\kelvin}$ the bulk carriers do not play any essential role in transport and points to formation of inversion/accumulation quasi-2D layers at the interfaces of the film, or, alternatively, to formation of topological surface states with similar properties in samples with inverted band order. Note that in both cases the bulk holes play an important role in maintaining overall charge neutrality in the sample.

\section{Methods}

\begin{figure}[tb]
	\centering
	\includegraphics[width=\linewidth]{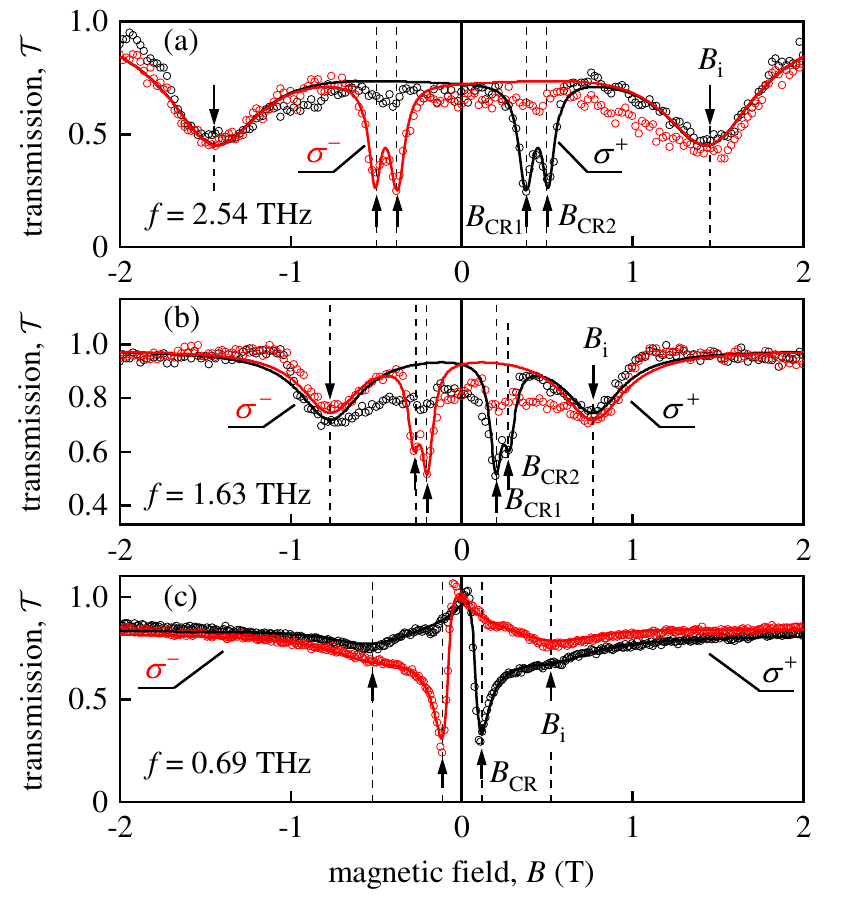}
	\caption{Magnetotransmission normalized to its maximal value for sample \#A with inverted band order in the flat region. Measurements were done in Faraday geometry [see figure~\ref{fig2}~(a)] at $T=\SI{4.2}{\kelvin}$ using $\sigma^+$ and $\sigma^-$ polarized radiation at three different frequencies: $f=\SI{2.54}{\tera\hertz}$ [panel (a)], $f=\SI{1.63}{\tera\hertz}$ [panel (b)], and $f=\SI{0.69}{\tera\hertz}$  [panel (c)].  In all panels, the experimental data points are shown by circles, whereas the solid lines present the fits described in the text. In all cases, we observe sharp deep minima at positions of cyclotron resonances [$B=\pm B_\mathrm{CR 1}$ and $B=\pm B_\mathrm{CR 2}$ in panels (a) and (b), and $B=\pm B_\mathrm{CR}$ in panel (c)], see arrows marked at the positive $B$ side. These CRs show up at either positive or negative $B$ depending on the helicity of incoming radiation. At higher $B$, we observe symmetric wide minima centered at $B=\pm B_\mathrm{i}$, which are not sensitive to the radiation helicity. The resonance field values and corresponding cyclotron masses are collected in Table \ref{BCRtable}.
}
	\label{fig4}
\end{figure}

\noindent
For optical excitation we use linearly polarized radiation from a continuous wave molecular gas laser operating at frequencies $f=\SI{2.54}{\tera\hertz}$ (wavelength $\lambda=\SI{118}{\micro\metre}$, photon energy $\hbar\omega=\SI{10.5}{\milli\electronvolt}$), $\SI{1.63}{\tera\hertz}$ ($\lambda=\SI{184}{\micro\metre}$, $\hbar\omega=\SI{6.74}{\milli\electronvolt}$), and $\SI{0.69}{\tera\hertz}$ ($\lambda=\SI{432}{\micro\metre}$, $\hbar\omega=\SI{2.87}{\milli\electronvolt}$)~\cite{Kvon2008a,Olbrich2011}. The laser provided radiation power ranging between 40 and $\SI{120}{\milli\watt}$ depending on the wavelength.
The radiation was modulated by an optical chopper operating at a frequency of about 80~Hz.
The samples were placed in a temperature controllable optical cryostat with $z$-cut quartz and TPX (4-methyl-1-pentene) windows.
In order to block visible and near-infrared radiation, the windows were additionally covered by black polyethylene foil.
The laser beam was focused using off-axis parabolic mirrors and controlled by a pyroelectric camera~\cite{Ganichev2002a}.  It had an almost Gaussian shape with the spot diameter ranging from 1.5 to $\SI{3}{\milli\metre}$ depending on the wavelength.
Room temperature lambda plates made of $x\mathrm{-cut}$ quartz were used to modify the state of polarization, in particular, to obtain circularly polarized right-handed ($\sigma^{+}$) and left-handed ($\sigma^{-}$) radiation. For time resolved measurements we additionally used a pulsed optically pumped THz laser with a pulse duration of about 100~ns, a repetition rate of 1~Hz, and a peak power $P$ of up to 60~kW~\cite{Ganichev2002b,Weber2008,Drexler2012}. The laser emitted frequency lines in the range between 0.6 and 2.6~THz. 

A split coil superconductive magnet was used to study magnetic field dependencies of radiation transmission and photocurrents.
Most of experiments were performed with normally incident THz radiation and magnetic field oriented either perpendicular to the film surface (Faraday geometry) or in-plane (Voigt geometry), see figures~\ref{fig2} (a) and  \ref{fig2} (b), respectively.
Additionally, experiments in Faraday geometry with the sample tilted by an angle $\theta$ were carried out, see figure~\ref{fig2} (c).

For experiments on radiation transmission a pyroelectric detector was placed behind the sample (see figure~\ref{fig2}). Photovoltage and photoconductivity measurements were carried out using a standard lock-in technique. Photovoltage was measured using the phase-locked voltage drop across the sample under the chopper-modulated THz radiation in the absence of external dc bias. 
To measure the photoconductivity, an external dc bias was applied and photoresponses for positive and negative bias polarities were subtracted in order to exclude photogalvanic contributions and extract the photoconductivity signal.
Alternatively, in some photoconductivity measurements we applied a low-frequency ac bias and used a double-modulation technique described in Refs.~\cite{Otteneder2018,Kozlov2011}.

\begin{table*}[tb]
	\centering
\begin{tabular*}{\textwidth}{c@{\extracolsep{\fill}} c c c c c c c c c}
	\hline
	\hline
	sample & Cd content $x$ & \makecell{frequency\\ $f (\SI{}{\tera\hertz})$} & \makecell{$B_\mathrm{CR1} (\SI{}{\tesla})$ \\ / $B_\mathrm{CR} (\SI{}{\tesla})$} & $B_\mathrm{CR2}(\SI{}{\tesla})$ & $B_\mathrm{i}(\SI{}{\tesla})$ &  \makecell{$B_\mathrm{CR1}^\text{V} (\SI{}{\tesla})$ \\ / $B_\mathrm{CR}^\text{V} (\SI{}{\tesla})$} & $B_\mathrm{CR2}^\text{V}(\SI{}{\tesla})$ & $m_\mathrm{CR1}(m_e)$ & $m_\mathrm{CR2}(m_e)$ \\ 
	\hline
	\hline 
	\rule{0pt}{3ex} 
	\# A & 0.15 
	& 2.54 & 0.38 & 0.50 & 1.45 & & & 0.00417 & 0.00551\\
	& & 1.63 & 0.20 & 0.27 & 0.77 & 0.16 & 0.24 & 0.00348 & 0.00465\\
	& & 0.69 & 0.09$^\ast$ & 0.11 & 0.52$^\ast$ & & & 0.00352 & 0.00443  \\
	
	\#B & 0.15 
	& 2.54 & 0.33 & 0.49 & 1.25 & 0.36 & 0.50 & 0.00365 & 0.00535\\
	& & 1.63 & 0.20 & 0.28 & 0.92 & 0.20 & 0.28 & 0.00339 & 0.00474\\
	
	\# C & 0.15 
	& 2.54 & 0.30 & 0.46 & 1.35$^\ast$ & 0.46 & -- & 0.00328 & 0.00512 \\

	\# D & 0.22 
	& 2.54 & 0.81 & -- & 2.33$^\ast$ & 0.82 & -- & 0.008927 & -- \\
	& & 1.63 & 0.49 & -- & 1.45$^\ast$ & 0.45 & -- & 0.008415 & -- \\

	\# E & 0.18 
	& 2.54 & 0.28 & -- & 0.77$^\ast$ & & & 0.003086 & -- \\
	& & 0.69 & 0.065 & -- & 0.25$^\ast$ & & & 0.002626 & -- \\
	\hline
	\hline
\end{tabular*}
\caption{Magnetic field positions of resonances and corresponding cyclotron masses extracted from magnetotransmission and photovoltage traces obtained at $T=\SI{4.2}{\kelvin}$. Note that resonances marked with an asterisk were visible in photovoltage only, and were not resolved in magnetotransmission.}
\label{BCRtable}
\end{table*}

\section{Results}
\label{results}

\subsection{Cyclotron resonances in \texorpdfstring{\ce{Cd_{0.15}Hg_{0.85}Te}}{CdHgTe} films with inverted band order and smooth interfaces}
\label{secAB}

\begin{figure}[tb]
	\centering
	\includegraphics[width=\linewidth]{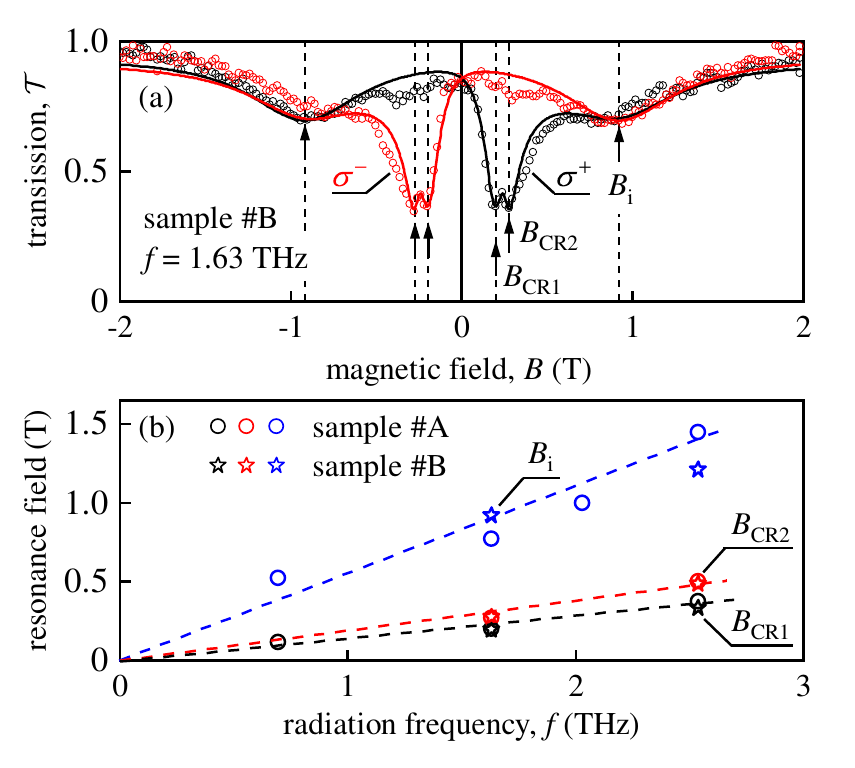}
	\caption{(a) Normalized magnetotransmission of  $f=\SI{1.63}{\tera\hertz}$ radiation measured on sample \#B in Faraday geometry at $T=\SI{4.2}{\kelvin}$. Two well resolved CR dips emerge at either $B=B_\mathrm{CR 1,\,2}$ or $B=-B_\mathrm{CR 1,\,2}$ depending on the helicity ($\sigma^+$ or $\sigma^-$) of incoming radiation. Additional helicity insensitive resonances are detected at $B=\pm B_\mathrm{i}$ in both traces. 
	Solid lines are fits according to Eq.~\eqref{transmissionformula} including additional lorentzians at $\pm B_\mathrm{i}$. (b) Magnetic field positions of the observed resonances $B_\mathrm{CR 1}$, $B_\mathrm{CR 2}$, and $B_\mathrm{i}$ (as marked)
	plotted against radiation frequency for sample \#A (circles) and \#B (stars) (see also Table \ref{BCRtable}). Dashed lines present linear fits to the data.}
	\label{fig5}
\end{figure}

\noindent
We begin with the presentation of results of magnetotransmission and photovoltage measurements performed on almost identical \texorpdfstring{\ce{Cd_{0.15}Hg_{0.85}Te}}{CdHgTe} films \#A and \#B, see figures~\ref{fig1} (b) and  \ref{fig1} (c). The flat layer in both samples has Cd content $x\simeq 0.15$, corresponding to inverted band order for temperatures below $T\simeq90$~K \cite{Teppe2016}. On both sides of the flat layer the Cd content gradually increases providing smooth interfaces to regions with normal band order. Magnetotransmission data recorded in Faraday geometry for samples \#A and \#B are presented, correspondingly, in figures~\ref{fig4} and \ref{fig5}. The data here and below generally manifest three resonant dips for both negative and positive values of the out-of-plane magnetic field $B$. The positions of observed resonances at $B=\pm B_\mathrm{CR 1,2}$, $B=\pm B_\mathrm{CR}$, and $B=\pm B_\mathrm{i}$, marked by arrows in figures~\ref{fig4} and \ref{fig5}, are always symmetrically offset from the origin $B=0$. 

The main focus of this work are sharp dips at low $B$ which, as shown below, originate from the CR of conduction electrons. By contrast, the broad minima at $B=\pm B_\mathrm{i}$ are attributed to photo-ionization of impurities and will be addressed separately in Sec.~\ref{impurity}. The data in figures~\ref{fig4} and \ref{fig5} (a) demonstrate that the low-field resonances are sensitive to the radiation helicity. While for the right-handed circular polarization $\sigma^+$ they appear only at positive $B$, for the left-handed circular polarization $\sigma^-$ they are present at negative $B$ only. The position of these resonances scales linearly with the radiation frequency, see figure~\ref{fig5} (b). All these features indicate that the low-$B$ sharp dips are indeed caused by the CR of negatively charged free carriers. 

More specifically, magnetotransmission traces in figure~\ref{fig4}~(a) that were recorded at the highest radiation frequency $f=\SI{2.54}{\tera\hertz}$ show two distinct sharp dips at $B=\pm B_\mathrm{CR 1}$ and $B=\pm B_\mathrm{CR 2}$, which clearly reveals the presence of two kinds of electrons with different cyclotron masses (the corresponding values are collected in Table \ref{BCRtable}). It is worth mentioning at this point that the corresponding cyclotron masses below $\text{0.01} m_0$ may be attributed both to bulk electrons and electrons in topological surface states, which are expected to exist in the films with inverted band order.  At lower frequency $f=\SI{1.64}{\tera\hertz}$ the CR dips start to merge, and become indistinguishable at the lowest frequency $f=\SI{0.69}{\tera\hertz}$ where a single merged CR dip $B_\mathrm{CR}$ is observed, see figures~\ref{fig4}~(b) and \ref{fig4}~(c). The positions of the resolved CR dips for $f=\SI{2.54}{\tera\hertz}$ and $f=\SI{1.64}{\tera\hertz}$ are found to be almost identical for both samples \#A and \#B, see figures~\ref{fig4} and \ref{fig5} and Table \ref{BCRtable}.

Solid lines in figure~\ref{fig4} are magnetotransmission fits obtained using a two-component Drude model with account for interference effects due to multiple reflections in the substrate. Specifically, the complex THz conductivity $\sigma$ is modeled as a sum of contributions $\sigma_j$ from each electron transport channel $j$,  with individual $\sigma_j$ given by the Drude formula,  
\begin{equation}
\sigma_j=\dfrac{e n_j}{\mu_j^{-1}+i (B_{\text{CR}j}\pm B)}\;.
\end{equation}
Here $e$ is the elementary charge, $n_j$ and $\mu_j$ are the electron sheet density and mobility in the transport channel $j$, $B_{\text{CR}j}=m_j 2\pi f/e$ defines the position of CR for a given THz frequency $f$ through the cyclotron mass $m_j$, and $+$ ($-$) sign before $B$ corresponds to the left-handed (right-handed) circular polarization of the THz radiation. The fraction of transmitted power is then expressed through $\tilde{\sigma}=\sigma/2\epsilon_0 c$ as \cite{Abstreiter1976,Hermann2016}
\begin{equation}
\label{transmissionformula}
T(B)=\left|\left(1+\tilde\sigma\right)\cos\phi-i\frac{1+n_\text{r}^2+2\tilde{\sigma}}{2 n_\text{r}}\sin\phi\right|^{-2}\;.
\end{equation} 
Here $\phi$ is the interference phase accumulated after single reflection in the substrate, $n_\text{r}$ is the refractive index of the substrate, $\epsilon_0$ the permittivity of free space, and $c$ the speed of light. Using Eq.~(\ref{transmissionformula}) we were able to precisely fit all magnetotransmission traces. This demonstrates that the model based on Drude approximation is capable to reproduce the experimental observations. At the same time, the magnetotransmission traces do not provide sufficient data to convincingly extract all relevant parameters of the model. The fits to experimental data required an additional constant offset accounting for non-resonant contributions of other transport channels. Two identical lorentzians centered at $B=\pm B_\mathrm{i}$ were included into the model to account for the impurity resonances discussed in Sec.~\ref{impurity}.

\begin{figure}[tb]
	\centering
	\includegraphics[width=\linewidth]{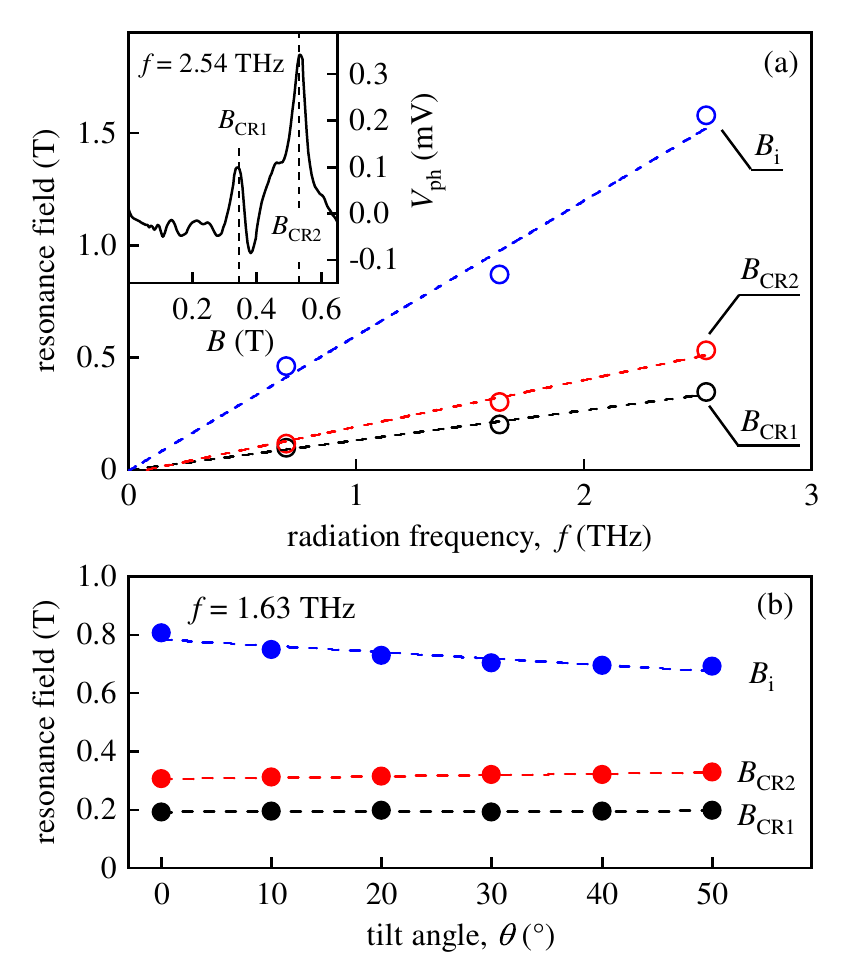}
	\caption{(a) Resonance field values $B_\mathrm{CR 1}$, $B_\mathrm{CR 2}$, and $B_\mathrm{i}$ (as marked) extracted from photovoltage measurements on sample \#A at $T=\SI{4.2}{\kelvin}$ in Faraday configuration. The low-$B$ part of a typical photovoltage trace, obtained at radiation frequency $f=\SI{2.54}{\tera\hertz}$, is shown in the inset.  (b) Dependence of the resonance field values on the tilt angle obtained from photovoltage measurements on sample \#A at $T=\SI{4.2}{\kelvin}$ under $f=\SI{1.63}{\tera\hertz}$ radiation in the Faraday configuration with tilted sample, see figure~\ref{fig2} (c). Dashed lines are linear fits to the data.}
	\label{fig6}
\end{figure}

The cyclotron resonances were also detected in photovoltage experiments. A typical photovoltage trace, shown in inset to figure~\ref{fig6} (a), features two peaks coinciding with the position of the CR dips in magnetotransmission as indicated by vertical dashed lines. Similar to the transmission dips, the position of the peaks scales linearly with the radiation frequency, see figure~\ref{fig6} (a). Note that in the photovoltage response two CR peaks are clearly resolved even for the lowest frequency $f=\SI{0.69}{\tera\hertz}$, in which case the CR dips in magnetotransmission get merged, see figure~\ref{fig4} (c).

\begin{figure}[tb]
	\centering
	\includegraphics[width=\linewidth]{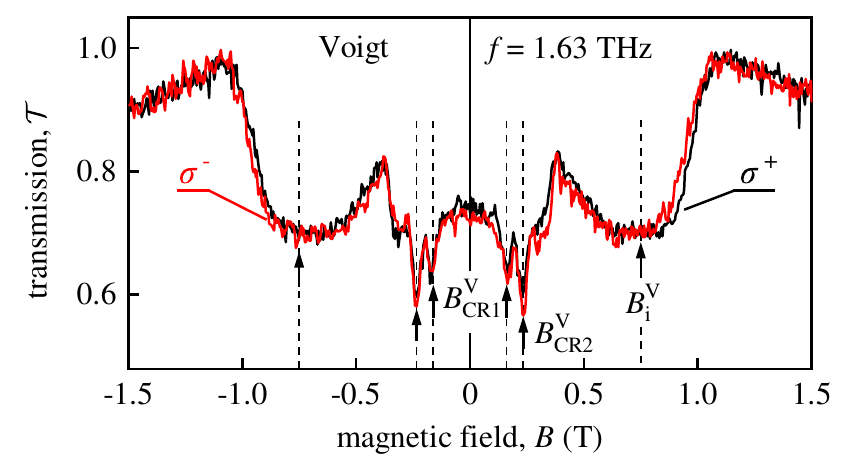}
	\caption{Normalized magnetotransmission of $f=\SI{1.63}{\tera\hertz}$ radiation measured in Voigt configuration on sample \#A  at $T=\SI{4.2}{\kelvin}$. Two pairs of well resolved CR dips emerge symmetrically at $B=B^\mathrm{V}_\mathrm{CR 1,\,2}$ and $B=-B^\mathrm{V}_\mathrm{CR 1,\,2}$ for both circular polarizations $\sigma^+$ (black curve) and $\sigma^-$ (red curve). Additional broad resonances are detected at $B=\pm B_\mathrm{i}$ in both traces. All resonance field values are collected in Table \ref{BCRtable}.
	}
	\label{fig7}
\end{figure}

Strikingly, applying a tilted magnetic field [see the experimental setup in figure~\ref{fig2} (c)] we obtain that the CR positions are independent of the tilt angle $\theta$, see figure~\ref{fig6}~(b). This observation demonstrates that the resonances are excited in a 3D electron gas. Measurements of the magnetotransmission in Voigt geometry [figure~\ref{fig2} (b)] confirm this observation showing that even for magnetic field oriented in-plane both CRs are still present. Note that, as expected, in Voigt geometry the CR dips become helicity-independent and show up symmetrically for two polarities of magnetic field, see figure~\ref{fig7}.

\subsection{Cyclotron resonances in \texorpdfstring{\ce{Cd_{0.15}Hg_{0.85}Te}}{CdHgTe} film with inverted band order and sharp top interface}
\label{secC}

For sample \#C the behavior of resonances becomes qualitatively different despite the fact that this sample has almost the same design and that the Cd concentration in the flat region remains the same as in samples \#A and \#B. The only difference is that instead of smooth top interface as in samples \#A and \#B, sample \#C has an abrupt boundary between the \texorpdfstring{\ce{Cd_{0.15}Hg_{0.85}Te}}{CdHgTe} film and \texorpdfstring{\ce{Cd_{0.85}Hg_{0.15}Te}}{CdHgTe} cap layer, see figure~\ref{fig1} (d). 

\begin{figure}[tb]
	\centering
	\includegraphics[width=\linewidth]{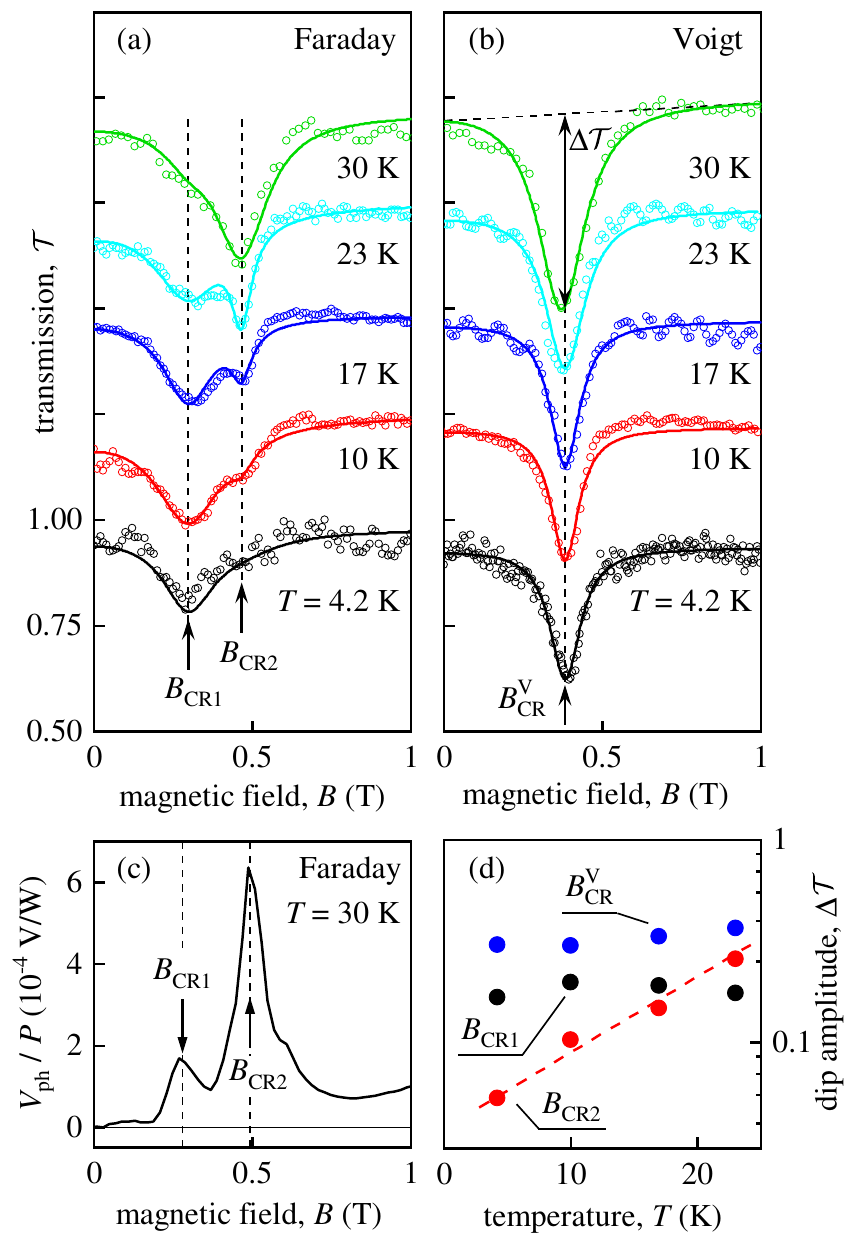}
	\caption{Normalized magnetotransmission of $f=\SI{2.54}{\tera\hertz}$ radiation obtained on sample \#C at different temperatures (as marked) in Faraday (a) and Voigt (b) configurations. Traces are vertically offset by 0.25 for better visibility.	Solid lines are fits using Eq.~\eqref{transmissionformula}. (c) Photovoltage measured on sample \#C in Faraday configuration at $T=\SI{30}{\kelvin}$. (d) Temperature dependence of the amplitudes $\Delta \mathcal{T}$ of the CR dips at $B_\text{CR1,\,2}$ and $B_\text{CR}^\text{V}$ (as marked) extracted from the magnetotransmission data presented in panels (a) and (b).
The dashed line shows a linear fit for the amplitude of CR dip at $B_\text{CR2}$. The corresponding CR field values are collected in Table \ref{BCRtable}.}
	\label{fig8}
\end{figure}

Similar to samples \#A and \#B, in Faraday geometry sample \#C clearly manifests two CRs in both magnetotransmission [see, e.g., blue curve in figure~\ref{fig8} (a)] and photovoltage [figure \ref{fig8} (c)]. However, in sharp contrast to samples with two smooth interfaces, in Voigt configuration only one CR dip 
remains, see figure \ref{fig8} (b). 
The fact that one of the resonances disappears in Voigt geometry provides a clear evidence that it originates from 2D confined electron states.
Figure \ref{fig8} (d) shows the temperature dependence of the dip amplitude $\Delta \mathcal{T}$. For resonances at $B_\text{CR1}$ (Faraday geometry) and $B_\text{CR}^\text{V}$ (Voigt configuration) we observe that $\Delta \mathcal{T}$ is almost independent of temperature.
By contrast, the strength of the CR resonance at $B_\text{CR2}$, which is present in Faraday geometry only, strongly increases with rising temperature. 
Such temperature evolution is strikingly similar to that found recently for topological surface states in strained 80 nm and 200 nm HgTe films \cite{Dantscher2015,Candussio2019}.
Taken together, the above observations clearly indicate formation of 2D surface states at the abrupt interface between materials with inverted and normal band order.

\subsection{Cyclotron resonances in films with normal band order}

We proceed to results obtained on films \#D (with $x=0.22$) and \#E ($x=0.18$) which are expected to have normal band order. Indeed, the data we are going to discuss were obtained at liquid helium temperature at which the critical Cd concentration marking the transition from normal to inverted band order is $x_c\simeq 0.17$ \cite{Teppe2016,Rigaux1980}. Thus, in both samples the Cd concentration in the flat region is higher than $x_c$ meaning that they should behave as conventional narrow-gap semiconductors.

\begin{figure}[tb]
	\centering
	\includegraphics[width=\linewidth]{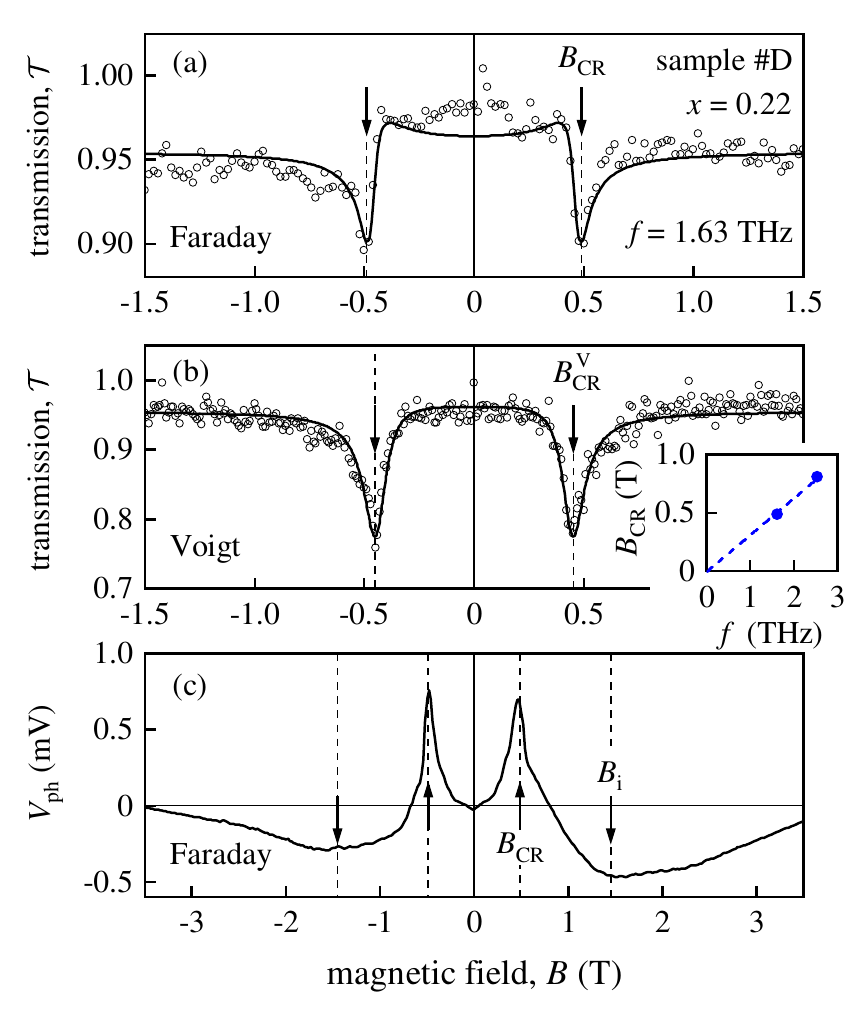}
	\caption{Normalized magnetotransmission of  $f=\SI{1.63}{\tera\hertz}$ linearly polarized radiation recorded on sample \#D with normal band order ($x=0.22$) at $T=\SI{4.2}{\kelvin}$ in Faraday (a) and Voigt (b) configurations. Solid lines are fits using Eq.~\eqref{transmissionformula}. (c) Photovoltage measured under the same conditions in Faraday configuration. Inset shows the CR field extracted from data obtained in Faraday configuration for two frequencies. All resonance field values are collected in Table \ref{BCRtable}.
	}
	\label{fig9}
\end{figure}

\begin{figure}[tb]
	\centering
	\includegraphics[width=\linewidth]{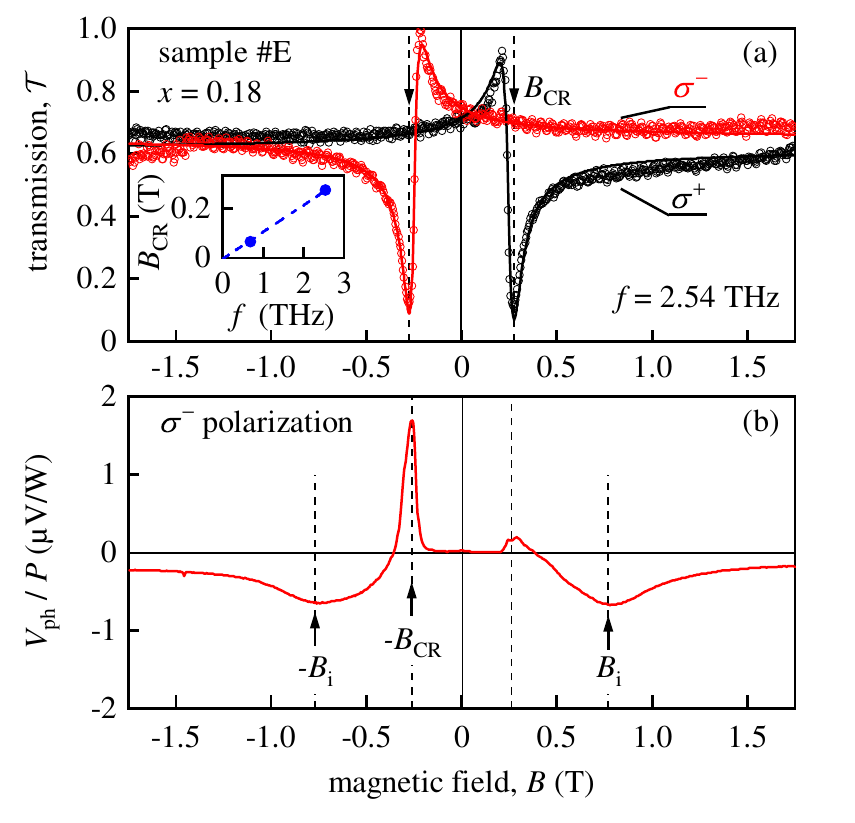}
	\caption{(a) Magnetotransmission of circularly polarized $f=\SI{2.54}{\tera\hertz}$ radiation measured on sample \#E with $x=0.18$ in Faraday geometry at $T=\SI{4.2}{\kelvin}$. The CR fields extracted from magnetotransmission data display a linear frequency scaling shown in inset. (b) Photovoltage measured under the same conditions for $\sigma^{-}$ polarization. The extracted resonance field values are collected in Table \ref{BCRtable}. }
	\label{fig10}
\end{figure}

In contrast to previously discussed materials with $x=0.15$, for these samples we observe a single CR in both Faraday and Voigt geometries. Magnetotransmission data for samples \#D and \#E is shown in figure~\ref{fig9}~(a), \ref{fig9}~(b), and figure~\ref{fig10}~(a). The data in figure~\ref{fig9} (a) and \ref{fig9}~(b) were obtained with linearly polarized THz radiation. Correspondingly, the CR dip is present for both polarities of magnetic fields. For circularly polarized radiation the CR dip appears either in positive ($\sigma^+$) or negative ($\sigma^-$) $B$, see figure \ref{fig10} (a). 

It is worth mentioning that the CR positions and the corresponding cyclotron masses obtained on samples \#D and \#E  (see Table \ref{BCRtable} and insets in figure~\ref{fig9} and \ref{fig10}) are substantially different from those obtained on other samples with $x=0.15$. This is consistent with our expectations: the largest cyclotron mass is obtained for sample \#D with $x=0.22$ which is supposed to have the largest band gap, while the lowest mass is obtained for sample \#D with $x=0.18$, closest to the critical Cd concentration $x_c\simeq 0.17$ corresponding to a gapless material with a linear spectrum. We also mention that a strong asymmetry of the CR feature in figure \ref{fig10} (a) is well captured by our fitting model and manifests significant role of interference effects included into Eq.~(\ref{transmissionformula}). Indeed, according to this formula, a symmetric shape of the transmission dip is expected either in conditions of constructive interference ($\sin \phi=0$) or destructive interference ($\cos \phi=0$). For intermediate values of $\phi$ the shape can be highly asymmetric \cite{Abstreiter1976}, and observation of such strong asymmetry is a clear indication of the importance of interference effects.

The CR resonances were also detected in the photovoltage response. Under linearly polarized radiation as in figure~\ref{fig9} (c), two symmetric CR peaks emerge in positive and negative $B$, whereas  the circularly polarized radiation in figure~\ref{fig10} (b) gives rise to a single pronounced CR peak for the corresponding active magnetic field polarity only. In this particular case, an additional peak with substantially smaller magnitude was detected for CR passive magnetic field $B>0$, most probably caused by parasitic antenna effect produced by contact wires.

\subsection{Impurity-related resonances}
\label{impurity}

\begin{figure}[tb]
	\centering
	\includegraphics[width=\linewidth]{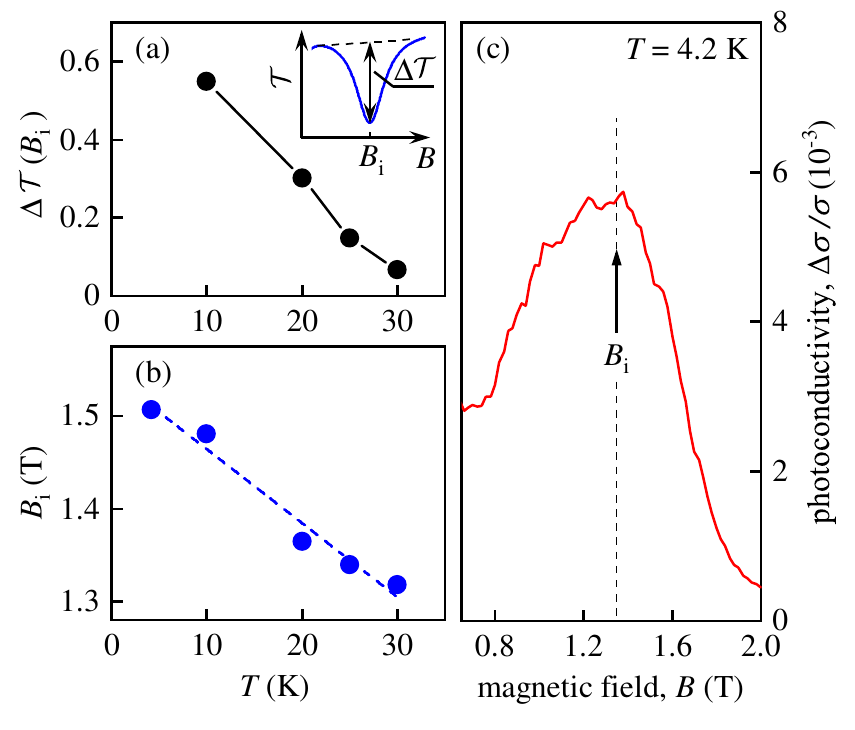}
	\caption{(a) Temperature dependence of the amplitude of lorentzian dips in magnetotransmission around the impurity resonance at $B=B_\text{i}$, extracted from data measured on sample \#A at $f=\SI{2.54}{\tera\hertz}$ as illustrated in the inset. (b) Temperature dependence of the impurity resonance position $B_\text{i}$ extracted from the same data. (c) Photoconductivity $\Delta\sigma$ normalized to the dark dc conductivity $\sigma$ showing a resonant peak at position $B=B_\mathrm{i}$ of the impurity resonance. These data were obtained at $T=\SI{4.2}{\kelvin}$ on sample \#A under $f=\SI{2.54}{\tera\hertz}$ radiation with applied current of $\SI{100}{\micro\ampere}$.}
	\label{fig11}
\end{figure}

\noindent Apart from the sharp cyclotron resonances at low $|B|$ which constitute the main subject of this work, all samples manifested broad resonant features at higher magnetic fields which, as discussed below, can be naturally attributed to photo-ionization of impurities and thus are referred here as impurity resonances. The main properties of the impurity resonances, observed in magnetotransmission as well as in photovoltage and photoconductivity, can be summarized as follows. First of all, impurity resonances appear symmetrically with respect to $B=0$ at $B=\pm B_i$ in both Faraday and Voigt geometries, and are insensitive to helicity of radiation (see, e.g., figure~\ref{fig4}). This fact alone provides a clear evidence that they may not be caused by the CR. The resonance positions $B=\pm B_i$ scale linearly with radiation frequency, as exemplified for samples \#A and \#B in figure \ref{fig5} (b). In contrast to the CR, in which case the amplitude of dips in transmission is either independent of temperature or increases with rising $T$, the amplitude of dips at $B=\pm B_i$ decreases quickly with growing $T$ and vanishes at $T\approx \SI{30}{\kelvin}$, see figure \ref{fig11} (a). The position $B_\text{i}$ of impurity resonances shifts to lower $|B|$ as $T$ grows, see figure~\ref{fig11} (b), whereas the CR positions are $T$-independent. The photoconductivity signal, shown in figure \ref{fig11} (c), significantly increases around the position of the impurity resonance.
The position of impurity resonances for a given radiation frequency is substantially different for samples with different cadmium concentration $x$, see Table~\ref{BCRtable}. In samples with $x=0.15$ in the flat region (inverted band order) the impurity resonances were detected both in transmission and photovoltage, while in samples with $x>x_c$ (normal band order) they were present in photovoltage but not visible in magnetotransmission.

On a qualitative level, the above features are consistent with the photo-ionization mechanism if one takes into account that the activation energy of shallow impurities can significantly increase in magnetic field, the effect well-known as magnetic freeze-out of impurities \cite{Dyakonov1969}. In this scenario, in the limit of low $B$ the activation energy is so small that all impurities are thermally ionized making the photo-ionization processes impossible \cite{Goldman1986}. In stronger $|B|$, the activation energy of impurities increases such that at sufficiently low $T$ the majority of impurities are in the neutral state and, and the same time, the photon energy is high enough to promote electrons into the unbound states in the conduction band. Under the same conditions, the total occupation of the conduction states is lowered making the relative contribution of the photo-excited carriers more pronounced. This is the parametric region of the impurity resonance where one expects most remarkable influence of the photo-ionization processes on magnetotransmission and photovoltage signals. In still stronger $|B|$ and low $T$, the photo-ionization is blocked as soon as the energy distance between the ground level of impurity and available unbound states exceeds the photon energy. Photo-ionization is thus possible in an intermediate range of $|B|$ centered at a certain $B_i$, and should be effective at sufficiently low temperatures only. Furthermore, the position of $B_i$ should increase with the radiation frequency, and should also be highly sensitive to temperature and to the Cd content $x$, as both of them modify the spectrum and, therefore, the position of impurity levels. On top of that, the temperature dependence can be affected by processes of photo-excitation between the hydrogen-like impurity levels with subsequent thermal excitation into the continuum of unbound conduction state~\cite{Lifshitz1965,Gershenzon1973}.

\begin{figure}[tb]
	\centering
	\includegraphics[width=\linewidth]{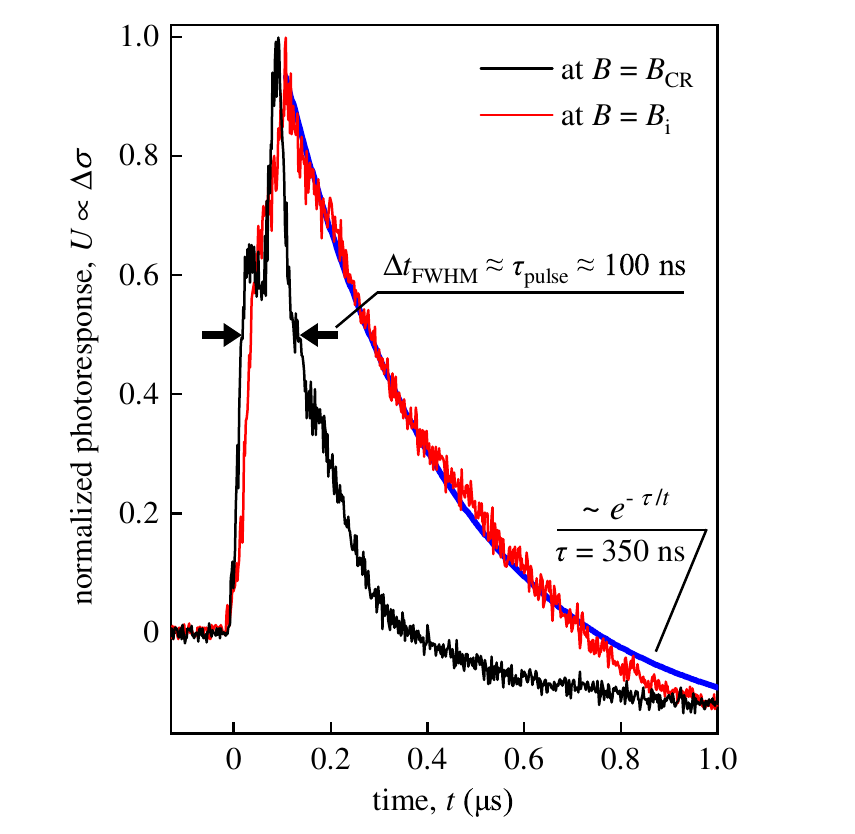}
	\caption{Time-resolved photoconductivity response $U\propto \Delta\sigma$ (normalized to its maximum value) obtained on sample \#A at liquid helium temperature using a pulsed THz laser operating at $f=\SI{2.03}{\tera\hertz}$. The black curve demonstrates a fast photoresponse at magnetic field $B_\mathrm{CR}=\SI{0.29}{\tesla}$ corresponding to the cyclotron resonance. The full width at half maximum of the CR photoresponse, $\Delta t_\text{FWHM}\simeq 100$~ns,  coincides with the corresponding width $\tau_\text{pulse}$ of the laser pulse profile (not shown). By contrast, the photoresponse recorded at magnetic field $B_\mathrm{i}=\SI{1}{\tesla}$, corresponding to the impurity resonance (red curve), shows a much slower decay which can be well fitted by an exponential function $\exp(-t/\tau)$ (blue line) with a decay time $\tau=\SI{350}{\nano\second}$.}
	\label{fig12}
\end{figure}

Assuming that the resonance at $B_\text{i}$ is related to ionization of impurity states, it is also natural to expect different kinetics of the photoresponse with respect to that at CR. These expectations were confirmed by time-resolved photoconductivity measurements performed using a pulsed THz laser. Figure \ref{fig12} shows that indeed the decay of the photoresponse at position $B=B_\text{i}$ of the impurity resonance is several times longer than that measured at position $B=B_\text{CR}$ of the CR. In the latter case (black trace in figure~\ref{fig12}), the photo-signal essentially reproduces the form of the THz laser pulse (not shown), meaning that kinetics of free carriers responsible for the CR is fast on the time scale of pulse duration, given by the full width at half maximum $\tau_\text{pulse}\simeq\SI{100}{\nano\second}$. By contrast, the photosignal measured at $B=B_\text{i}$ (red trace in figure~\ref{fig12}) features a significant long-time tail, which can be well fitted by an exponential $\exp(-t/\tau)$ (blue line in figure~\ref{fig12}) with a decay time $\tau=\SI{350}{\nano\second}$.

\section{Discussion}
\label{discussion}

We now return to the main observation: a qualitatively different behavior of one of CRs in samples with sharp and smooth interfaces between materials with inverted and normal band order, see Sec.~\ref{secAB} and Sec.~\ref{secC}. The fact that one of the CRs disappears in the Voigt geometry for sample \#C with a sharp interface is a clear indicator of a true 2D nature of carriers causing this resonance. The emergence of at least one surface state is indeed guaranteed by the topologically non-trivial transition from inverted to normal band order at the interface. In Faraday geometry, the same resonance exhibits a strong temperature dependence, which additionally distinguishes it from other resonances, which are present in both Voigt and Faraday geometries and presumably originate from carriers localized at smooth interfaces. These results indicate that smooth interfaces lead to emergence of multiple surface states, known as Volkov-Pankratov states (VPS)~\cite{Inhofer2017,Mahler2019,Volkov1985,Tchoumakov2017}. As the number $N$ of these states is not fixed by topology (apart from the requirement $N\geq 1$), their number and properties should be sensitive to a particular choice of materials, growth conditions, form of interfaces, strain profile etc. The dynamics of carriers occupying multiple VPS states can be effectively three-dimensional, similar to conventional quasi-2D electrons occupying multiple subbands in a wide quantum well, which would immediately explain our observations. Here we present a general discussion and model calculations supporting this interpretation of the distinct behaviors at smooth and sharp interfaces.

The surface states in systems with sharp interfaces are well studied for strained HgTe films \cite{Bruene2011,Dantscher2015,Hancock2011,Kibis2019}. These topologically protected states are caused by the inversion of $\Gamma_{6}$ and $\Gamma_{8}$ bands at the interface. The energy of the surface states depends of the wave vector $\bm{k}_{\|}$ in the interface plane and lies in the interval between the corresponding energy levels of the light and heavy holes \cite{Bruene2011,Pankratov1987}. If the crystal is strained, there is a gap between the light- and heavy-hole subbands; the energy of the lowest $k_{\|}=0$ topological surface state coincides with top of the heavy-hole subband, while in the limit $|k_{\|}|\rightarrow \infty$ the dispersion of the high-energy surface states approaches that of the light holes. As a result, ideally the low-temperature transport properties of a strained HgTe film with Fermi energy lying in the gap between the light- and heavy-hole subbands should be solely determined by the topologically protected surface states, since all bulk states in this case are gapped. Here it is important to mention that, despite the very existence of a topological surface state stems from the inversion of light-hole band and conduction band at the interface, the dispersion of the surface states is strongly affected by hybridization with the heavy holes. The CdHgTe films with an inverted band order share the above qualitative properties of strained HgTe films, but usually feature much smaller band gaps which are controlled by the Cd content of the alloy.

\begin{figure}
\centering
\includegraphics[width=\linewidth]{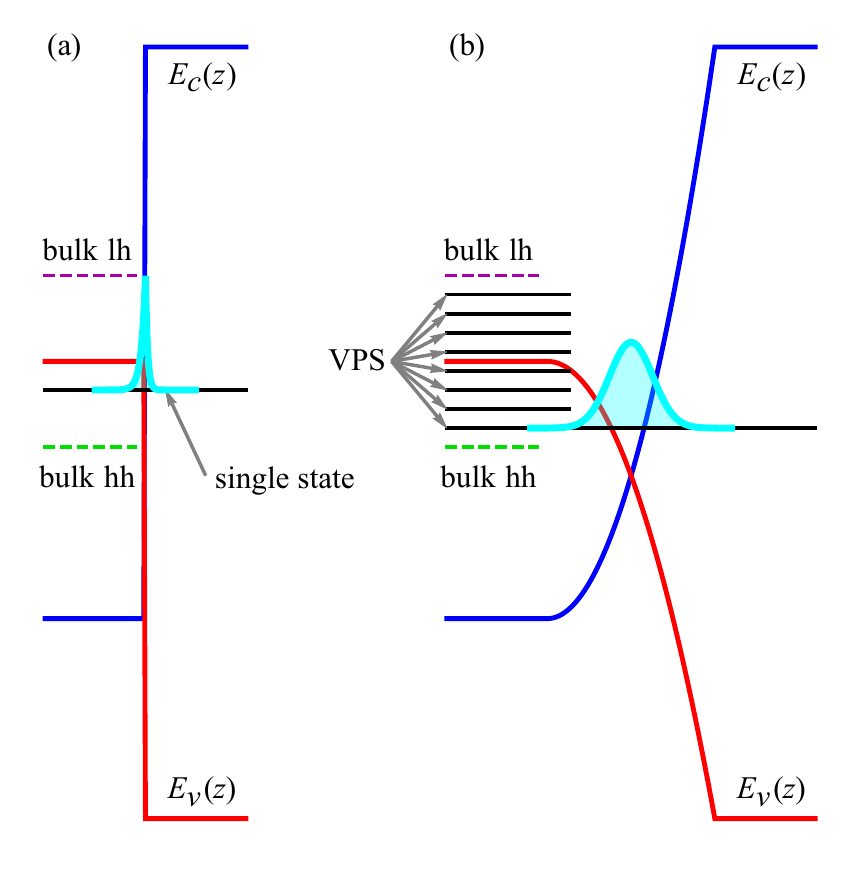}
\caption{Illustration of the edge states forming at sharp [panel (a)]
and smooth [panel (b)] interfaces. Light blue lines with filled area below
them show the calculated probability density distributions for the
edge states in both cases. 
The sharp interface hosts a single well-localized topologically
protected state, while the smooth interface gives rise to multiple
Volkov-Pankratov states (VPS), which are weakly localized in space.
Red and blue lines show the real space profiles $E_v(z)$ and $E_c(z)$
of the corresponding band edges. The edge states emerge in the region
between energies of the bulk light and heavy holes, which are shown by
green and magenta dashed lines and are marked as ``bulk lh'' and ``bulk
hh'', correspondingly).
}
\label{theoryfig}
\end{figure}

For a smooth interface with a band inversion the situation is different. The calculations below show that instead of a single topologically protected state, interface now can host multiple VPS states \cite{Inhofer2017,Mahler2019,Volkov1985,Tchoumakov2017}. Similar to the case of sharp interface, the energies of VPS lie in between the energies of the strain-split light and heavy holes. Figure~\ref{theoryfig} schematically shows a sharp interface [as in sample \#C, panel (a)] and a smooth interface [as in sample \#A or \#B, panel (b)]. The left side in panels (a) and (b) presents the flat region of \texorpdfstring{\ce{Cd$_{0.15}$Hg$_{0.85}$Te}}{CdHgTe},  while the left side shows transition to the top cap layer with normal band order and a much wider band gap. For the sharp interface, the spatial extension of the topologically protected surface state is determined by an exponential decay of the wave function inside the film and the barrier. On the contrary, the extension of VPS localized at a smooth interface is governed by the spatial gradient of the band gap near the band inversion point. The spatial distribution of the probability densities for the corresponding wave functions is illustrated in Fig.~\ref{theoryfig}. If one increases the width of the transition region between the topological insulator and normal insulator, both the total number of the VPS and their widths increase. This explains why two samples \#A and \#C with an almost identical design, but different interface structure, show different behavior in the Voigt configuration. Sample \#C hosts 2D topologically protected states which do not experience cyclotron resonance in an in-plane of the interface magnetic field, while sample \#A exhibits behavior typical for quasi-2D system with many occupied subbands.

For numerical calculations we implement the 6-band Kane model taking into account a static strain of the crystal which causes a splitting of the heavy and light-hole subbands. In the basis of the $\Gamma_6$ and $\Gamma_8$ states, the Kane Hamiltonian has the form~\cite{Bir1974}
\begin{equation}
                \label{KaneH}
                H =
                \begin{pmatrix}
                   E_c  I_{2} &  H_{\text{cv}}\\[0.1cm]
                    H_{\text{cv}}^\dagger & E_v I_4+H_\text{BP}
                \end{pmatrix}\,,
\end{equation}
where  $I_{n}$ is the $n\times n$ identity matrix,
\begin{equation}
   H_{\text{cv}}^\dag = P_\text{cv}
   \begin{pmatrix}
                    -\dfrac{k_{x}-i k_{y}}{\sqrt{2}}  & 0 \\[0.4cm]
                    \sqrt{\dfrac{2}{3}} k_{z} & -\dfrac{k_{x}-i
k_{y}}{\sqrt{6}} \\[0.4cm]
                    \dfrac{k_{x}+ i k_{y}}{\sqrt{6}} &
\sqrt{\dfrac{2}{3}} k_{z}  \\[0.3cm]
                    0 & \dfrac{k_{x}+ i k_{y}}{\sqrt{2}}
    \end{pmatrix} \,.
\end{equation}
Here $P_\text{cv}$ is the Kane parameter and $H_\text{BP}$ is the Bir-Pikus Hamiltonian. In order to phenomenologically model the gap between light and heavy holes we introduce a non-zero $u_{zz}$ component of strain tensor, in which case the Bir-Pikus Hamiltonian has the form $H_\text{BP}=b u_{zz} \operatorname{diag}(-1,1,1,-1)$, where $b$ is the valence band deformation potential and $\operatorname{diag}$ stands for a diagonal matrix. Below we set $u_{zz}$ such that the heavy-hole subband is below the light-hole subband. This Hamiltonian for the bulk CdHgTe has three double-degenerate eigenstates: the light-hole subband states with dispersion $\varepsilon_l(k)$ on top, the dispersionless states in the heavy-hole subband $\varepsilon_h(k)$ below them, and the low-lying conduction band states $\varepsilon_c(k)$, explicitly given by
\begin{align*}
&\varepsilon_{l,c}(k)=E_0\pm\sqrt{\delta^2+2P_\text{cv}^2
k^2/3}\:,\\
&\varepsilon_h(k)=E_v-b u_{zz}\:,
\end{align*}
with $2E_0=E_v+b u_{zz}+E_c$ and $2 \delta=E_v+b u_{zz}-E_c$. 

To calculate the VPS spectrum and wave functions for a smooth interface we take into account the dependence of $E_c$ and $E_v$ on the coordinate. Using the parameters $2 m_0 (P_\text{cv}/\hbar)^2=18.8$~eV~\cite{Novik2005}, the band gap $E_c-E_v=-32$~meV for Cd$_{0.15}$Hg$_{0.85}$Te \cite{Laurenti1990}, and the dependencies of $E_c(z)$ and $E_v(z)$ found from the profile of compound distribution $x$ shown in Fig.~\ref{fig1} (a), we calculated VPS for sample \#A. In this calculation, we set $b u_{zz}=2$~meV and add a static electric field $E=2$~kV/cm at the band closing. At $k_{\|}=0.02$~nm$^{-1}$, such calculation yields three VPS states with energies $3$, $5$ and $8$~meV with respect to the heavy-hole energy at $\Gamma$ point in Cd$_{0.15}$Hg$_{0.85}$Te film. The probability density of the VPS is distributed over hundreds of nanometers, which is greater or of the order of the cyclotron radius $r_\text{c}$ of charge carriers in the bulk CdHgTe film. The latter can be found as $r_c=v/\omega$ using $\omega=2\pi f$ and an estimate $v=\sqrt{2/3}P_\text{cv}/\hbar\simeq 10^8$~cm/sec for the carrier velocity. This yields values of $r_c=65$~nm, 100~nm, and 240~nm for $f=2.54$~THz, $1.63$~THz, and $0.69$~THz, respectively. We also find that the number of VPS emerging at such smooth interface is highly sensitive to changes of the band gap, as well as to the energy splitting between light- and heavy-holes caused by deformation and static electric field. The latter parameters are not known precisely and are introduced here as phenomenological parameters. Taken together, the presented calculations confirm that smooth interfaces in our samples can indeed host many VPS leading to a quasi-2D behavior of the CRs, which are thus present for both in-plane and out-of-plane orientations of the magnetic field.

\section{Summary} 
\label{summary}

\noindent
To summarize, the presented low-tem\-perature THz spectroscopy studies of cyclotron resonances in thick films of \texorpdfstring{\ce{Cd_{x}Hg_{1-x}Te}}{CdHgTe}  unambiguously show that two-dimensional surface states are present only in structures with a sharp interface separating the internal layer having an inverted band order ($x<0.17$) and a cap layer having normal band order. By contrast, conventional smooth interfaces from inverted to normal band order, involving regions of \texorpdfstring{\ce{Cd_{x}Hg_{1-x}Te}}{CdHgTe} with gradually changing Cd content $x$, make the surface states effectively three-dimensional such that the corresponding cyclotron resonances are clearly visible even for an in-plane orientation of applied magnetic field. Our observations, supported by calculations describing formation of the surface states in films with sharp and smooth interfaces, clearly demonstrate that future studies of topologically protected surface states of \texorpdfstring{\ce{Cd_{x}Hg_{1-x}Te}}{CdHgTe} films require structures with sharp interfaces. Apart from the cyclotron resonances, all samples manifest broad helicity-independent resonances which emerge at higher magnetic fields and show a slower kinetics and higher temperature sensitivity. These broad resonances are attributed to photo-ionization and magnetic freeze-out of impurity states.

\section*{acknowledgement}
\label{acknow}
We thank S.A.Tarasenko for fruitful discussions.
The support from the DFG priority program SPP 1666 ``Topological Insulators: Materials - Fundamental Properties - Devices'' (project GA501/12-2) and the IRAP programme of the Foundation for Polish Science (grant MAB/2018/9, project CENTERA) and TEAM project POIR.04.04.00-00-3D76/16 (TEAM/2016-3/25) are gratefully acknowledged. 
V.V.B. and S.A.D. acknowledge support by the Volkswagen Stiftung (Az. 97738).
I.D. acknowledges support from the Deutsche Forschungsgemeinschaft (project no. DM1/4-1).
G.V.B. acknowledges support from the Russian Science Foundation (project no. 17-12-01265) and ``BASIS'' foundation.
I.Y. acknowledges support by the National Science Centre, Poland (grant No. UMO-2017/25/N/ST3/00408).

\bibliography{resonances_CdHgTe_main_file_arxiv_2020.01.13}

\begin{thebibliography}{63}%
\makeatletter
\providecommand \@ifxundefined [1]{%
 \@ifx{#1\undefined}
}%
\providecommand \@ifnum [1]{%
 \ifnum #1\expandafter \@firstoftwo
 \else \expandafter \@secondoftwo
 \fi
}%
\providecommand \@ifx [1]{%
 \ifx #1\expandafter \@firstoftwo
 \else \expandafter \@secondoftwo
 \fi
}%
\providecommand \natexlab [1]{#1}%
\providecommand \enquote  [1]{``#1''}%
\providecommand \bibnamefont  [1]{#1}%
\providecommand \bibfnamefont [1]{#1}%
\providecommand \citenamefont [1]{#1}%
\providecommand \href@noop [0]{\@secondoftwo}%
\providecommand \href [0]{\begingroup \@sanitize@url \@href}%
\providecommand \@href[1]{\@@startlink{#1}\@@href}%
\providecommand \@@href[1]{\endgroup#1\@@endlink}%
\providecommand \@sanitize@url [0]{\catcode `\\12\catcode `\$12\catcode
  `\&12\catcode `\#12\catcode `\^12\catcode `\_12\catcode `\%12\relax}%
\providecommand \@@startlink[1]{}%
\providecommand \@@endlink[0]{}%
\providecommand \url  [0]{\begingroup\@sanitize@url \@url }%
\providecommand \@url [1]{\endgroup\@href {#1}{\urlprefix }}%
\providecommand \urlprefix  [0]{URL }%
\providecommand \Eprint [0]{\href }%
\providecommand \doibase [0]{http://dx.doi.org/}%
\providecommand \selectlanguage [0]{\@gobble}%
\providecommand \bibinfo  [0]{\@secondoftwo}%
\providecommand \bibfield  [0]{\@secondoftwo}%
\providecommand \translation [1]{[#1]}%
\providecommand \BibitemOpen [0]{}%
\providecommand \bibitemStop [0]{}%
\providecommand \bibitemNoStop [0]{.\EOS\space}%
\providecommand \EOS [0]{\spacefactor3000\relax}%
\providecommand \BibitemShut  [1]{\csname bibitem#1\endcsname}%
\let\auto@bib@innerbib\@empty
\bibitem [{\citenamefont {Bernevig}\ \emph {et~al.}(2006)\citenamefont
  {Bernevig}, \citenamefont {Hughes},\ and\ \citenamefont
  {Zhang}}]{Bernevig2006}%
  \BibitemOpen
  \bibfield  {author} {\bibinfo {author} {\bibfnamefont {B.~A.}\ \bibnamefont
  {Bernevig}}, \bibinfo {author} {\bibfnamefont {T.~L.}\ \bibnamefont
  {Hughes}}, \ and\ \bibinfo {author} {\bibfnamefont {S.-C.}\ \bibnamefont
  {Zhang}},\ }\href {\doibase 10.1126/science.1133734} {\bibfield  {journal}
  {\bibinfo  {journal} {Science}\ }\textbf {\bibinfo {volume} {314}},\ \bibinfo
  {pages} {1757} (\bibinfo {year} {2006})}\BibitemShut {NoStop}%
\bibitem [{\citenamefont {König}\ \emph {et~al.}(2008)\citenamefont {König},
  \citenamefont {Buhmann}, \citenamefont {Molenkamp}, \citenamefont {Hughes},
  \citenamefont {Liu}, \citenamefont {Qi},\ and\ \citenamefont
  {Zhang}}]{Koenig2008}%
  \BibitemOpen
  \bibfield  {author} {\bibinfo {author} {\bibfnamefont {M.}~\bibnamefont
  {König}}, \bibinfo {author} {\bibfnamefont {H.}~\bibnamefont {Buhmann}},
  \bibinfo {author} {\bibfnamefont {L.~W.}\ \bibnamefont {Molenkamp}}, \bibinfo
  {author} {\bibfnamefont {T.}~\bibnamefont {Hughes}}, \bibinfo {author}
  {\bibfnamefont {C.-X.}\ \bibnamefont {Liu}}, \bibinfo {author} {\bibfnamefont
  {X.-L.}\ \bibnamefont {Qi}}, \ and\ \bibinfo {author} {\bibfnamefont {S.-C.}\
  \bibnamefont {Zhang}},\ }\href {\doibase 10.1143/jpsj.77.031007} {\bibfield
  {journal} {\bibinfo  {journal} {J. Phys. Soc. Jpn.}\ }\textbf {\bibinfo
  {volume} {77}},\ \bibinfo {pages} {031007} (\bibinfo {year}
  {2008})}\BibitemShut {NoStop}%
\bibitem [{\citenamefont {Hasan}\ and\ \citenamefont {Kane}(2010)}]{Hasan2010}%
  \BibitemOpen
  \bibfield  {author} {\bibinfo {author} {\bibfnamefont {M.~Z.}\ \bibnamefont
  {Hasan}}\ and\ \bibinfo {author} {\bibfnamefont {C.~L.}\ \bibnamefont
  {Kane}},\ }\href {\doibase 10.1103/revmodphys.82.3045} {\bibfield  {journal}
  {\bibinfo  {journal} {Rev. Mod. Phys.}\ }\textbf {\bibinfo {volume} {82}},\
  \bibinfo {pages} {3045} (\bibinfo {year} {2010})}\BibitemShut {NoStop}%
\bibitem [{\citenamefont {Moore}(2010)}]{Moore2010}%
  \BibitemOpen
  \bibfield  {author} {\bibinfo {author} {\bibfnamefont {J.~E.}\ \bibnamefont
  {Moore}},\ }\href {\doibase 10.1038/nature08916} {\bibfield  {journal}
  {\bibinfo  {journal} {Nature}\ }\textbf {\bibinfo {volume} {464}},\ \bibinfo
  {pages} {194} (\bibinfo {year} {2010})}\BibitemShut {NoStop}%
\bibitem [{\citenamefont {Qi}\ and\ \citenamefont {Zhang}(2011)}]{Qi2011}%
  \BibitemOpen
  \bibfield  {author} {\bibinfo {author} {\bibfnamefont {X.-L.}\ \bibnamefont
  {Qi}}\ and\ \bibinfo {author} {\bibfnamefont {S.-C.}\ \bibnamefont {Zhang}},\
  }\href {\doibase 10.1103/revmodphys.83.1057} {\bibfield  {journal} {\bibinfo
  {journal} {Rev. Mod. Phys.}\ }\textbf {\bibinfo {volume} {83}},\ \bibinfo
  {pages} {1057} (\bibinfo {year} {2011})}\BibitemShut {NoStop}%
\bibitem [{\citenamefont {Ortmann}\ \emph {et~al.}(2015)\citenamefont
  {Ortmann}, \citenamefont {Roche},\ and\ \citenamefont
  {Valenzuela}}]{Ortmann2015}%
  \BibitemOpen
  \bibinfo {editor} {\bibfnamefont {F.}~\bibnamefont {Ortmann}}, \bibinfo
  {editor} {\bibfnamefont {S.}~\bibnamefont {Roche}}, \ and\ \bibinfo {editor}
  {\bibfnamefont {S.~O.}\ \bibnamefont {Valenzuela}},\ eds.,\ \href
  {https://www.ebook.de/de/product/22845571/topological_insulators.html} {\emph
  {\bibinfo {title} {Topological Insulators: Fundamentals and Perspectives}}}\
  (\bibinfo  {publisher} {Wiley VCH Verlag GmbH},\ \bibinfo {year}
  {2015})\BibitemShut {NoStop}%
\bibitem [{\citenamefont {Maier}\ \emph {et~al.}(2012)\citenamefont {Maier},
  \citenamefont {Oostinga}, \citenamefont {Knott}, \citenamefont {Brüne},
  \citenamefont {Virtanen}, \citenamefont {Tkachov}, \citenamefont
  {Hankiewicz}, \citenamefont {Gould}, \citenamefont {Buhmann},\ and\
  \citenamefont {Molenkamp}}]{Maier2012}%
  \BibitemOpen
  \bibfield  {author} {\bibinfo {author} {\bibfnamefont {L.}~\bibnamefont
  {Maier}}, \bibinfo {author} {\bibfnamefont {J.~B.}\ \bibnamefont {Oostinga}},
  \bibinfo {author} {\bibfnamefont {D.}~\bibnamefont {Knott}}, \bibinfo
  {author} {\bibfnamefont {C.}~\bibnamefont {Brüne}}, \bibinfo {author}
  {\bibfnamefont {P.}~\bibnamefont {Virtanen}}, \bibinfo {author}
  {\bibfnamefont {G.}~\bibnamefont {Tkachov}}, \bibinfo {author} {\bibfnamefont
  {E.~M.}\ \bibnamefont {Hankiewicz}}, \bibinfo {author} {\bibfnamefont
  {C.}~\bibnamefont {Gould}}, \bibinfo {author} {\bibfnamefont
  {H.}~\bibnamefont {Buhmann}}, \ and\ \bibinfo {author} {\bibfnamefont
  {L.~W.}\ \bibnamefont {Molenkamp}},\ }\href {\doibase
  10.1103/physrevlett.109.186806} {\bibfield  {journal} {\bibinfo  {journal}
  {Phys. Rev. Lett.}\ }\textbf {\bibinfo {volume} {109}},\ \bibinfo {pages}
  {186806} (\bibinfo {year} {2012})}\BibitemShut {NoStop}%
\bibitem [{\citenamefont {Ren}\ \emph {et~al.}(2019)\citenamefont {Ren},
  \citenamefont {Pientka}, \citenamefont {Hart}, \citenamefont {Pierce},
  \citenamefont {Kosowsky}, \citenamefont {Lunczer}, \citenamefont {Schlereth},
  \citenamefont {Scharf}, \citenamefont {Hankiewicz}, \citenamefont
  {Molenkamp}, \citenamefont {Halperin},\ and\ \citenamefont
  {Yacoby}}]{Ren2019}%
  \BibitemOpen
  \bibfield  {author} {\bibinfo {author} {\bibfnamefont {H.}~\bibnamefont
  {Ren}}, \bibinfo {author} {\bibfnamefont {F.}~\bibnamefont {Pientka}},
  \bibinfo {author} {\bibfnamefont {S.}~\bibnamefont {Hart}}, \bibinfo {author}
  {\bibfnamefont {A.~T.}\ \bibnamefont {Pierce}}, \bibinfo {author}
  {\bibfnamefont {M.}~\bibnamefont {Kosowsky}}, \bibinfo {author}
  {\bibfnamefont {L.}~\bibnamefont {Lunczer}}, \bibinfo {author} {\bibfnamefont
  {R.}~\bibnamefont {Schlereth}}, \bibinfo {author} {\bibfnamefont
  {B.}~\bibnamefont {Scharf}}, \bibinfo {author} {\bibfnamefont {E.~M.}\
  \bibnamefont {Hankiewicz}}, \bibinfo {author} {\bibfnamefont {L.~W.}\
  \bibnamefont {Molenkamp}}, \bibinfo {author} {\bibfnamefont {B.~I.}\
  \bibnamefont {Halperin}}, \ and\ \bibinfo {author} {\bibfnamefont
  {A.}~\bibnamefont {Yacoby}},\ }\href {\doibase 10.1038/s41586-019-1148-9}
  {\bibfield  {journal} {\bibinfo  {journal} {Nature}\ }\textbf {\bibinfo
  {volume} {569}},\ \bibinfo {pages} {93} (\bibinfo {year} {2019})}\BibitemShut
  {NoStop}%
\bibitem [{\citenamefont {Brüne}\ \emph {et~al.}(2011)\citenamefont {Brüne},
  \citenamefont {Liu}, \citenamefont {Novik}, \citenamefont {Hankiewicz},
  \citenamefont {Buhmann}, \citenamefont {Chen}, \citenamefont {Qi},
  \citenamefont {Shen}, \citenamefont {Zhang},\ and\ \citenamefont
  {Molenkamp}}]{Bruene2011}%
  \BibitemOpen
  \bibfield  {author} {\bibinfo {author} {\bibfnamefont {C.}~\bibnamefont
  {Brüne}}, \bibinfo {author} {\bibfnamefont {C.~X.}\ \bibnamefont {Liu}},
  \bibinfo {author} {\bibfnamefont {E.~G.}\ \bibnamefont {Novik}}, \bibinfo
  {author} {\bibfnamefont {E.~M.}\ \bibnamefont {Hankiewicz}}, \bibinfo
  {author} {\bibfnamefont {H.}~\bibnamefont {Buhmann}}, \bibinfo {author}
  {\bibfnamefont {Y.~L.}\ \bibnamefont {Chen}}, \bibinfo {author}
  {\bibfnamefont {X.~L.}\ \bibnamefont {Qi}}, \bibinfo {author} {\bibfnamefont
  {Z.~X.}\ \bibnamefont {Shen}}, \bibinfo {author} {\bibfnamefont {S.~C.}\
  \bibnamefont {Zhang}}, \ and\ \bibinfo {author} {\bibfnamefont {L.~W.}\
  \bibnamefont {Molenkamp}},\ }\href {\doibase 10.1103/physrevlett.106.126803}
  {\bibfield  {journal} {\bibinfo  {journal} {Phys. Rev. Lett.}\ }\textbf
  {\bibinfo {volume} {106}},\ \bibinfo {pages} {126803} (\bibinfo {year}
  {2011})}\BibitemShut {NoStop}%
\bibitem [{\citenamefont {Nowack}\ \emph {et~al.}(2013)\citenamefont {Nowack},
  \citenamefont {Spanton}, \citenamefont {Baenninger}, \citenamefont {König},
  \citenamefont {Kirtley}, \citenamefont {Kalisky}, \citenamefont {Ames},
  \citenamefont {Leubner}, \citenamefont {Brüne}, \citenamefont {Buhmann},
  \citenamefont {Molenkamp}, \citenamefont {Goldhaber-Gordon},\ and\
  \citenamefont {Moler}}]{Nowack2013}%
  \BibitemOpen
  \bibfield  {author} {\bibinfo {author} {\bibfnamefont {K.~C.}\ \bibnamefont
  {Nowack}}, \bibinfo {author} {\bibfnamefont {E.~M.}\ \bibnamefont {Spanton}},
  \bibinfo {author} {\bibfnamefont {M.}~\bibnamefont {Baenninger}}, \bibinfo
  {author} {\bibfnamefont {M.}~\bibnamefont {König}}, \bibinfo {author}
  {\bibfnamefont {J.~R.}\ \bibnamefont {Kirtley}}, \bibinfo {author}
  {\bibfnamefont {B.}~\bibnamefont {Kalisky}}, \bibinfo {author} {\bibfnamefont
  {C.}~\bibnamefont {Ames}}, \bibinfo {author} {\bibfnamefont {P.}~\bibnamefont
  {Leubner}}, \bibinfo {author} {\bibfnamefont {C.}~\bibnamefont {Brüne}},
  \bibinfo {author} {\bibfnamefont {H.}~\bibnamefont {Buhmann}}, \bibinfo
  {author} {\bibfnamefont {L.~W.}\ \bibnamefont {Molenkamp}}, \bibinfo {author}
  {\bibfnamefont {D.}~\bibnamefont {Goldhaber-Gordon}}, \ and\ \bibinfo
  {author} {\bibfnamefont {K.~A.}\ \bibnamefont {Moler}},\ }\href {\doibase
  10.1038/nmat3682} {\bibfield  {journal} {\bibinfo  {journal} {Nat. Mater.}\
  }\textbf {\bibinfo {volume} {12}},\ \bibinfo {pages} {787} (\bibinfo {year}
  {2013})}\BibitemShut {NoStop}%
\bibitem [{\citenamefont {Kozlov}\ \emph {et~al.}(2014)\citenamefont {Kozlov},
  \citenamefont {Kvon}, \citenamefont {Olshanetsky}, \citenamefont {Mikhailov},
  \citenamefont {Dvoretsky},\ and\ \citenamefont {Weiss}}]{Kozlov2014}%
  \BibitemOpen
  \bibfield  {author} {\bibinfo {author} {\bibfnamefont {D.~A.}\ \bibnamefont
  {Kozlov}}, \bibinfo {author} {\bibfnamefont {Z.~D.}\ \bibnamefont {Kvon}},
  \bibinfo {author} {\bibfnamefont {E.~B.}\ \bibnamefont {Olshanetsky}},
  \bibinfo {author} {\bibfnamefont {N.~N.}\ \bibnamefont {Mikhailov}}, \bibinfo
  {author} {\bibfnamefont {S.~A.}\ \bibnamefont {Dvoretsky}}, \ and\ \bibinfo
  {author} {\bibfnamefont {D.}~\bibnamefont {Weiss}},\ }\href {\doibase
  10.1103/PhysRevLett.112.196801} {\bibfield  {journal} {\bibinfo  {journal}
  {Phys. Rev. Lett.}\ }\textbf {\bibinfo {volume} {112}},\ \bibinfo {pages}
  {196801} (\bibinfo {year} {2014})}\BibitemShut {NoStop}%
\bibitem [{\citenamefont {Brüne}\ \emph {et~al.}(2014)\citenamefont {Brüne},
  \citenamefont {Thienel}, \citenamefont {Stuiber}, \citenamefont {Böttcher},
  \citenamefont {Buhmann}, \citenamefont {Novik}, \citenamefont {Liu},
  \citenamefont {Hankiewicz},\ and\ \citenamefont {Molenkamp}}]{Bruene2014}%
  \BibitemOpen
  \bibfield  {author} {\bibinfo {author} {\bibfnamefont {C.}~\bibnamefont
  {Brüne}}, \bibinfo {author} {\bibfnamefont {C.}~\bibnamefont {Thienel}},
  \bibinfo {author} {\bibfnamefont {M.}~\bibnamefont {Stuiber}}, \bibinfo
  {author} {\bibfnamefont {J.}~\bibnamefont {Böttcher}}, \bibinfo {author}
  {\bibfnamefont {H.}~\bibnamefont {Buhmann}}, \bibinfo {author} {\bibfnamefont
  {E.~G.}\ \bibnamefont {Novik}}, \bibinfo {author} {\bibfnamefont {C.-X.}\
  \bibnamefont {Liu}}, \bibinfo {author} {\bibfnamefont {E.~M.}\ \bibnamefont
  {Hankiewicz}}, \ and\ \bibinfo {author} {\bibfnamefont {L.~W.}\ \bibnamefont
  {Molenkamp}},\ }\href {\doibase 10.1103/physrevx.4.041045} {\bibfield
  {journal} {\bibinfo  {journal} {Phys. Rev. X}\ }\textbf {\bibinfo {volume}
  {4}},\ \bibinfo {pages} {041045} (\bibinfo {year} {2014})}\BibitemShut
  {NoStop}%
\bibitem [{\citenamefont {Olshanetsky}\ \emph {et~al.}(2015)\citenamefont
  {Olshanetsky}, \citenamefont {Kvon}, \citenamefont {Gusev}, \citenamefont
  {Levin}, \citenamefont {Raichev}, \citenamefont {Mikhailov},\ and\
  \citenamefont {Dvoretsky}}]{Olshanetsky2015}%
  \BibitemOpen
  \bibfield  {author} {\bibinfo {author} {\bibfnamefont {E.}~\bibnamefont
  {Olshanetsky}}, \bibinfo {author} {\bibfnamefont {Z.}~\bibnamefont {Kvon}},
  \bibinfo {author} {\bibfnamefont {G.}~\bibnamefont {Gusev}}, \bibinfo
  {author} {\bibfnamefont {A.}~\bibnamefont {Levin}}, \bibinfo {author}
  {\bibfnamefont {O.}~\bibnamefont {Raichev}}, \bibinfo {author} {\bibfnamefont
  {N.}~\bibnamefont {Mikhailov}}, \ and\ \bibinfo {author} {\bibfnamefont
  {S.}~\bibnamefont {Dvoretsky}},\ }\href {\doibase
  10.1103/physrevlett.114.126802} {\bibfield  {journal} {\bibinfo  {journal}
  {Phys. Rev. Lett.}\ }\textbf {\bibinfo {volume} {114}},\ \bibinfo {pages}
  {126802} (\bibinfo {year} {2015})}\BibitemShut {NoStop}%
\bibitem [{\citenamefont {Ma}\ \emph {et~al.}(2015)\citenamefont {Ma},
  \citenamefont {Calvo}, \citenamefont {Wang}, \citenamefont {Lian},
  \citenamefont {Mühlbauer}, \citenamefont {Brüne}, \citenamefont {Cui},
  \citenamefont {Lai}, \citenamefont {Kundhikanjana}, \citenamefont {Yang},
  \citenamefont {Baenninger}, \citenamefont {König}, \citenamefont {Ames},
  \citenamefont {Buhmann}, \citenamefont {Leubner}, \citenamefont {Molenkamp},
  \citenamefont {Zhang}, \citenamefont {Goldhaber-Gordon}, \citenamefont
  {Kelly},\ and\ \citenamefont {Shen}}]{Ma2015}%
  \BibitemOpen
  \bibfield  {author} {\bibinfo {author} {\bibfnamefont {E.~Y.}\ \bibnamefont
  {Ma}}, \bibinfo {author} {\bibfnamefont {M.~R.}\ \bibnamefont {Calvo}},
  \bibinfo {author} {\bibfnamefont {J.}~\bibnamefont {Wang}}, \bibinfo {author}
  {\bibfnamefont {B.}~\bibnamefont {Lian}}, \bibinfo {author} {\bibfnamefont
  {M.}~\bibnamefont {Mühlbauer}}, \bibinfo {author} {\bibfnamefont
  {C.}~\bibnamefont {Brüne}}, \bibinfo {author} {\bibfnamefont {Y.-T.}\
  \bibnamefont {Cui}}, \bibinfo {author} {\bibfnamefont {K.}~\bibnamefont
  {Lai}}, \bibinfo {author} {\bibfnamefont {W.}~\bibnamefont {Kundhikanjana}},
  \bibinfo {author} {\bibfnamefont {Y.}~\bibnamefont {Yang}}, \bibinfo {author}
  {\bibfnamefont {M.}~\bibnamefont {Baenninger}}, \bibinfo {author}
  {\bibfnamefont {M.}~\bibnamefont {König}}, \bibinfo {author} {\bibfnamefont
  {C.}~\bibnamefont {Ames}}, \bibinfo {author} {\bibfnamefont {H.}~\bibnamefont
  {Buhmann}}, \bibinfo {author} {\bibfnamefont {P.}~\bibnamefont {Leubner}},
  \bibinfo {author} {\bibfnamefont {L.~W.}\ \bibnamefont {Molenkamp}}, \bibinfo
  {author} {\bibfnamefont {S.-C.}\ \bibnamefont {Zhang}}, \bibinfo {author}
  {\bibfnamefont {D.}~\bibnamefont {Goldhaber-Gordon}}, \bibinfo {author}
  {\bibfnamefont {M.~A.}\ \bibnamefont {Kelly}}, \ and\ \bibinfo {author}
  {\bibfnamefont {Z.-X.}\ \bibnamefont {Shen}},\ }\href {\doibase
  10.1038/ncomms8252} {\bibfield  {journal} {\bibinfo  {journal} {Nat.
  Commun.}\ }\textbf {\bibinfo {volume} {6}},\ \bibinfo {pages} {7252}
  (\bibinfo {year} {2015})}\BibitemShut {NoStop}%
\bibitem [{\citenamefont {Kozlov}\ \emph {et~al.}(2016)\citenamefont {Kozlov},
  \citenamefont {Bauer}, \citenamefont {Ziegler}, \citenamefont {Fischer},
  \citenamefont {Savchenko}, \citenamefont {Kvon}, \citenamefont {Mikhailov},
  \citenamefont {Dvoretsky},\ and\ \citenamefont {Weiss}}]{Kozlov2016}%
  \BibitemOpen
  \bibfield  {author} {\bibinfo {author} {\bibfnamefont {D.}~\bibnamefont
  {Kozlov}}, \bibinfo {author} {\bibfnamefont {D.}~\bibnamefont {Bauer}},
  \bibinfo {author} {\bibfnamefont {J.}~\bibnamefont {Ziegler}}, \bibinfo
  {author} {\bibfnamefont {R.}~\bibnamefont {Fischer}}, \bibinfo {author}
  {\bibfnamefont {M.}~\bibnamefont {Savchenko}}, \bibinfo {author}
  {\bibfnamefont {Z.}~\bibnamefont {Kvon}}, \bibinfo {author} {\bibfnamefont
  {N.}~\bibnamefont {Mikhailov}}, \bibinfo {author} {\bibfnamefont
  {S.}~\bibnamefont {Dvoretsky}}, \ and\ \bibinfo {author} {\bibfnamefont
  {D.}~\bibnamefont {Weiss}},\ }\href {\doibase 10.1103/physrevlett.116.166802}
  {\bibfield  {journal} {\bibinfo  {journal} {Phys. Rev. Lett.}\ }\textbf
  {\bibinfo {volume} {116}},\ \bibinfo {pages} {166802} (\bibinfo {year}
  {2016})}\BibitemShut {NoStop}%
\bibitem [{\citenamefont {Inhofer}\ \emph {et~al.}(2017)\citenamefont
  {Inhofer}, \citenamefont {Tchoumakov}, \citenamefont {Assaf}, \citenamefont
  {F{\`{e}}ve}, \citenamefont {Berroir}, \citenamefont {Jouffrey},
  \citenamefont {Carpentier}, \citenamefont {Goerbig}, \citenamefont
  {Pla{\c{c}}ais}, \citenamefont {Bendias}, \citenamefont {Mahler},
  \citenamefont {Bocquillon}, \citenamefont {Schlereth}, \citenamefont
  {Brüne}, \citenamefont {Buhmann},\ and\ \citenamefont
  {Molenkamp}}]{Inhofer2017}%
  \BibitemOpen
  \bibfield  {author} {\bibinfo {author} {\bibfnamefont {A.}~\bibnamefont
  {Inhofer}}, \bibinfo {author} {\bibfnamefont {S.}~\bibnamefont {Tchoumakov}},
  \bibinfo {author} {\bibfnamefont {B.~A.}\ \bibnamefont {Assaf}}, \bibinfo
  {author} {\bibfnamefont {G.}~\bibnamefont {F{\`{e}}ve}}, \bibinfo {author}
  {\bibfnamefont {J.~M.}\ \bibnamefont {Berroir}}, \bibinfo {author}
  {\bibfnamefont {V.}~\bibnamefont {Jouffrey}}, \bibinfo {author}
  {\bibfnamefont {D.}~\bibnamefont {Carpentier}}, \bibinfo {author}
  {\bibfnamefont {M.~O.}\ \bibnamefont {Goerbig}}, \bibinfo {author}
  {\bibfnamefont {B.}~\bibnamefont {Pla{\c{c}}ais}}, \bibinfo {author}
  {\bibfnamefont {K.}~\bibnamefont {Bendias}}, \bibinfo {author} {\bibfnamefont
  {D.~M.}\ \bibnamefont {Mahler}}, \bibinfo {author} {\bibfnamefont
  {E.}~\bibnamefont {Bocquillon}}, \bibinfo {author} {\bibfnamefont
  {R.}~\bibnamefont {Schlereth}}, \bibinfo {author} {\bibfnamefont
  {C.}~\bibnamefont {Brüne}}, \bibinfo {author} {\bibfnamefont
  {H.}~\bibnamefont {Buhmann}}, \ and\ \bibinfo {author} {\bibfnamefont
  {L.~W.}\ \bibnamefont {Molenkamp}},\ }\href {\doibase
  10.1103/physrevb.96.195104} {\bibfield  {journal} {\bibinfo  {journal} {Phys.
  Rev. B}\ }\textbf {\bibinfo {volume} {96}},\ \bibinfo {pages} {195104}
  (\bibinfo {year} {2017})}\BibitemShut {NoStop}%
\bibitem [{\citenamefont {Maier}\ \emph {et~al.}(2017)\citenamefont {Maier},
  \citenamefont {Ziegler}, \citenamefont {Fischer}, \citenamefont {Kozlov},
  \citenamefont {Kvon}, \citenamefont {Mikhailov}, \citenamefont {Dvoretsky},\
  and\ \citenamefont {Weiss}}]{Maier2017}%
  \BibitemOpen
  \bibfield  {author} {\bibinfo {author} {\bibfnamefont {H.}~\bibnamefont
  {Maier}}, \bibinfo {author} {\bibfnamefont {J.}~\bibnamefont {Ziegler}},
  \bibinfo {author} {\bibfnamefont {R.}~\bibnamefont {Fischer}}, \bibinfo
  {author} {\bibfnamefont {D.}~\bibnamefont {Kozlov}}, \bibinfo {author}
  {\bibfnamefont {Z.~D.}\ \bibnamefont {Kvon}}, \bibinfo {author}
  {\bibfnamefont {N.}~\bibnamefont {Mikhailov}}, \bibinfo {author}
  {\bibfnamefont {S.~A.}\ \bibnamefont {Dvoretsky}}, \ and\ \bibinfo {author}
  {\bibfnamefont {D.}~\bibnamefont {Weiss}},\ }\href {\doibase
  10.1038/s41467-017-01684-0} {\bibfield  {journal} {\bibinfo  {journal} {Nat.
  Commun.}\ }\textbf {\bibinfo {volume} {8}},\ \bibinfo {pages} {2023}
  (\bibinfo {year} {2017})}\BibitemShut {NoStop}%
\bibitem [{\citenamefont {Imhof}\ \emph {et~al.}(2018)\citenamefont {Imhof},
  \citenamefont {Berger}, \citenamefont {Bayer}, \citenamefont {Brehm},
  \citenamefont {Molenkamp}, \citenamefont {Kiessling}, \citenamefont
  {Schindler}, \citenamefont {Lee}, \citenamefont {Greiter}, \citenamefont
  {Neupert},\ and\ \citenamefont {Thomale}}]{Imhof2018}%
  \BibitemOpen
  \bibfield  {author} {\bibinfo {author} {\bibfnamefont {S.}~\bibnamefont
  {Imhof}}, \bibinfo {author} {\bibfnamefont {C.}~\bibnamefont {Berger}},
  \bibinfo {author} {\bibfnamefont {F.}~\bibnamefont {Bayer}}, \bibinfo
  {author} {\bibfnamefont {J.}~\bibnamefont {Brehm}}, \bibinfo {author}
  {\bibfnamefont {L.~W.}\ \bibnamefont {Molenkamp}}, \bibinfo {author}
  {\bibfnamefont {T.}~\bibnamefont {Kiessling}}, \bibinfo {author}
  {\bibfnamefont {F.}~\bibnamefont {Schindler}}, \bibinfo {author}
  {\bibfnamefont {C.~H.}\ \bibnamefont {Lee}}, \bibinfo {author} {\bibfnamefont
  {M.}~\bibnamefont {Greiter}}, \bibinfo {author} {\bibfnamefont
  {T.}~\bibnamefont {Neupert}}, \ and\ \bibinfo {author} {\bibfnamefont
  {R.}~\bibnamefont {Thomale}},\ }\href {\doibase 10.1038/s41567-018-0246-1}
  {\bibfield  {journal} {\bibinfo  {journal} {Nat. Phys.}\ }\textbf {\bibinfo
  {volume} {14}},\ \bibinfo {pages} {925} (\bibinfo {year} {2018})}\BibitemShut
  {NoStop}%
\bibitem [{\citenamefont {Mahler}\ \emph {et~al.}(2019)\citenamefont {Mahler},
  \citenamefont {Mayer}, \citenamefont {Leubner}, \citenamefont {Lunczer},
  \citenamefont {Sante}, \citenamefont {Sangiovanni}, \citenamefont {Thomale},
  \citenamefont {Hankiewicz}, \citenamefont {Buhmann}, \citenamefont {Gould},\
  and\ \citenamefont {Molenkamp}}]{Mahler2019}%
  \BibitemOpen
  \bibfield  {author} {\bibinfo {author} {\bibfnamefont {D.~M.}\ \bibnamefont
  {Mahler}}, \bibinfo {author} {\bibfnamefont {J.-B.}\ \bibnamefont {Mayer}},
  \bibinfo {author} {\bibfnamefont {P.}~\bibnamefont {Leubner}}, \bibinfo
  {author} {\bibfnamefont {L.}~\bibnamefont {Lunczer}}, \bibinfo {author}
  {\bibfnamefont {D.~D.}\ \bibnamefont {Sante}}, \bibinfo {author}
  {\bibfnamefont {G.}~\bibnamefont {Sangiovanni}}, \bibinfo {author}
  {\bibfnamefont {R.}~\bibnamefont {Thomale}}, \bibinfo {author} {\bibfnamefont
  {E.~M.}\ \bibnamefont {Hankiewicz}}, \bibinfo {author} {\bibfnamefont
  {H.}~\bibnamefont {Buhmann}}, \bibinfo {author} {\bibfnamefont
  {C.}~\bibnamefont {Gould}}, \ and\ \bibinfo {author} {\bibfnamefont {L.~W.}\
  \bibnamefont {Molenkamp}},\ }\href {\doibase 10.1103/physrevx.9.031034}
  {\bibfield  {journal} {\bibinfo  {journal} {Phys. Rev. X}\ }\textbf {\bibinfo
  {volume} {9}},\ \bibinfo {pages} {031034} (\bibinfo {year}
  {2019})}\BibitemShut {NoStop}%
\bibitem [{\citenamefont {Kvon}\ \emph {et~al.}(2012)\citenamefont {Kvon},
  \citenamefont {Danilov}, \citenamefont {Kozlov}, \citenamefont {Zoth},
  \citenamefont {Mikhailov}, \citenamefont {Dvoretskii},\ and\ \citenamefont
  {Ganichev}}]{Kvon2012}%
  \BibitemOpen
  \bibfield  {author} {\bibinfo {author} {\bibfnamefont {Z.~D.}\ \bibnamefont
  {Kvon}}, \bibinfo {author} {\bibfnamefont {S.~N.}\ \bibnamefont {Danilov}},
  \bibinfo {author} {\bibfnamefont {D.~A.}\ \bibnamefont {Kozlov}}, \bibinfo
  {author} {\bibfnamefont {C.}~\bibnamefont {Zoth}}, \bibinfo {author}
  {\bibfnamefont {N.~N.}\ \bibnamefont {Mikhailov}}, \bibinfo {author}
  {\bibfnamefont {S.~A.}\ \bibnamefont {Dvoretskii}}, \ and\ \bibinfo {author}
  {\bibfnamefont {S.~D.}\ \bibnamefont {Ganichev}},\ }\href {\doibase
  10.1134/s002136401123007x} {\bibfield  {journal} {\bibinfo  {journal} {JETP
  Lett.}\ }\textbf {\bibinfo {volume} {94}},\ \bibinfo {pages} {816} (\bibinfo
  {year} {2012})}\BibitemShut {NoStop}%
\bibitem [{\citenamefont {Zholudev}\ \emph {et~al.}(2012)\citenamefont
  {Zholudev}, \citenamefont {Teppe}, \citenamefont {Orlita}, \citenamefont
  {Consejo}, \citenamefont {Torres}, \citenamefont {Dyakonova}, \citenamefont
  {Czapkiewicz}, \citenamefont {Wr{\'{o}}bel}, \citenamefont {Grabecki},
  \citenamefont {Mikhailov}, \citenamefont {Dvoretskii}, \citenamefont
  {Ikonnikov}, \citenamefont {Spirin}, \citenamefont {Aleshkin}, \citenamefont
  {Gavrilenko},\ and\ \citenamefont {Knap}}]{Zholudev2012}%
  \BibitemOpen
  \bibfield  {author} {\bibinfo {author} {\bibfnamefont {M.}~\bibnamefont
  {Zholudev}}, \bibinfo {author} {\bibfnamefont {F.}~\bibnamefont {Teppe}},
  \bibinfo {author} {\bibfnamefont {M.}~\bibnamefont {Orlita}}, \bibinfo
  {author} {\bibfnamefont {C.}~\bibnamefont {Consejo}}, \bibinfo {author}
  {\bibfnamefont {J.}~\bibnamefont {Torres}}, \bibinfo {author} {\bibfnamefont
  {N.}~\bibnamefont {Dyakonova}}, \bibinfo {author} {\bibfnamefont
  {M.}~\bibnamefont {Czapkiewicz}}, \bibinfo {author} {\bibfnamefont
  {J.}~\bibnamefont {Wr{\'{o}}bel}}, \bibinfo {author} {\bibfnamefont
  {G.}~\bibnamefont {Grabecki}}, \bibinfo {author} {\bibfnamefont
  {N.}~\bibnamefont {Mikhailov}}, \bibinfo {author} {\bibfnamefont
  {S.}~\bibnamefont {Dvoretskii}}, \bibinfo {author} {\bibfnamefont
  {A.}~\bibnamefont {Ikonnikov}}, \bibinfo {author} {\bibfnamefont
  {K.}~\bibnamefont {Spirin}}, \bibinfo {author} {\bibfnamefont
  {V.}~\bibnamefont {Aleshkin}}, \bibinfo {author} {\bibfnamefont
  {V.}~\bibnamefont {Gavrilenko}}, \ and\ \bibinfo {author} {\bibfnamefont
  {W.}~\bibnamefont {Knap}},\ }\href {\doibase 10.1103/physrevb.86.205420}
  {\bibfield  {journal} {\bibinfo  {journal} {Phys. Rev. B}\ }\textbf {\bibinfo
  {volume} {86}},\ \bibinfo {pages} {205420} (\bibinfo {year}
  {2012})}\BibitemShut {NoStop}%
\bibitem [{\citenamefont {Olbrich}\ \emph {et~al.}(2013)\citenamefont
  {Olbrich}, \citenamefont {Zoth}, \citenamefont {Vierling}, \citenamefont
  {Dantscher}, \citenamefont {Budkin}, \citenamefont {Tarasenko}, \citenamefont
  {Bel'kov}, \citenamefont {Kozlov}, \citenamefont {Kvon}, \citenamefont
  {Mikhailov}, \citenamefont {Dvoretsky},\ and\ \citenamefont
  {Ganichev}}]{Olbrich2013}%
  \BibitemOpen
  \bibfield  {author} {\bibinfo {author} {\bibfnamefont {P.}~\bibnamefont
  {Olbrich}}, \bibinfo {author} {\bibfnamefont {C.}~\bibnamefont {Zoth}},
  \bibinfo {author} {\bibfnamefont {P.}~\bibnamefont {Vierling}}, \bibinfo
  {author} {\bibfnamefont {K.-M.}\ \bibnamefont {Dantscher}}, \bibinfo {author}
  {\bibfnamefont {G.~V.}\ \bibnamefont {Budkin}}, \bibinfo {author}
  {\bibfnamefont {S.~A.}\ \bibnamefont {Tarasenko}}, \bibinfo {author}
  {\bibfnamefont {V.~V.}\ \bibnamefont {Bel'kov}}, \bibinfo {author}
  {\bibfnamefont {D.~A.}\ \bibnamefont {Kozlov}}, \bibinfo {author}
  {\bibfnamefont {Z.~D.}\ \bibnamefont {Kvon}}, \bibinfo {author}
  {\bibfnamefont {N.~N.}\ \bibnamefont {Mikhailov}}, \bibinfo {author}
  {\bibfnamefont {S.~A.}\ \bibnamefont {Dvoretsky}}, \ and\ \bibinfo {author}
  {\bibfnamefont {S.~D.}\ \bibnamefont {Ganichev}},\ }\href {\doibase
  10.1103/physrevb.87.235439} {\bibfield  {journal} {\bibinfo  {journal} {Phys.
  Rev. B}\ }\textbf {\bibinfo {volume} {87}},\ \bibinfo {pages} {235439}
  (\bibinfo {year} {2013})}\BibitemShut {NoStop}%
\bibitem [{\citenamefont {Shuvaev}\ \emph {et~al.}(2013)\citenamefont
  {Shuvaev}, \citenamefont {Pimenov}, \citenamefont {Astakhov}, \citenamefont
  {Mühlbauer}, \citenamefont {Brüne}, \citenamefont {Buhmann},\ and\
  \citenamefont {Molenkamp}}]{Shuvaev2013}%
  \BibitemOpen
  \bibfield  {author} {\bibinfo {author} {\bibfnamefont {A.}~\bibnamefont
  {Shuvaev}}, \bibinfo {author} {\bibfnamefont {A.}~\bibnamefont {Pimenov}},
  \bibinfo {author} {\bibfnamefont {G.~V.}\ \bibnamefont {Astakhov}}, \bibinfo
  {author} {\bibfnamefont {M.}~\bibnamefont {Mühlbauer}}, \bibinfo {author}
  {\bibfnamefont {C.}~\bibnamefont {Brüne}}, \bibinfo {author} {\bibfnamefont
  {H.}~\bibnamefont {Buhmann}}, \ and\ \bibinfo {author} {\bibfnamefont
  {L.~W.}\ \bibnamefont {Molenkamp}},\ }\href {\doibase 10.1063/1.4811496}
  {\bibfield  {journal} {\bibinfo  {journal} {Appl. Phys. Lett.}\ }\textbf
  {\bibinfo {volume} {102}},\ \bibinfo {pages} {241902} (\bibinfo {year}
  {2013})}\BibitemShut {NoStop}%
\bibitem [{\citenamefont {Pakmehr}\ \emph {et~al.}(2014)\citenamefont
  {Pakmehr}, \citenamefont {Bruene}, \citenamefont {Buhmann}, \citenamefont
  {Molenkamp}, \citenamefont {Stier},\ and\ \citenamefont
  {McCombe}}]{Pakmehr2014}%
  \BibitemOpen
  \bibfield  {author} {\bibinfo {author} {\bibfnamefont {M.}~\bibnamefont
  {Pakmehr}}, \bibinfo {author} {\bibfnamefont {C.}~\bibnamefont {Bruene}},
  \bibinfo {author} {\bibfnamefont {H.}~\bibnamefont {Buhmann}}, \bibinfo
  {author} {\bibfnamefont {L.~W.}\ \bibnamefont {Molenkamp}}, \bibinfo {author}
  {\bibfnamefont {A.~V.}\ \bibnamefont {Stier}}, \ and\ \bibinfo {author}
  {\bibfnamefont {B.~D.}\ \bibnamefont {McCombe}},\ }\href {\doibase
  10.1103/physrevb.90.235414} {\bibfield  {journal} {\bibinfo  {journal} {Phys.
  Rev. B}\ }\textbf {\bibinfo {volume} {90}},\ \bibinfo {pages} {235414}
  (\bibinfo {year} {2014})}\BibitemShut {NoStop}%
\bibitem [{\citenamefont {Dantscher}\ \emph {et~al.}(2015)\citenamefont
  {Dantscher}, \citenamefont {Kozlov}, \citenamefont {Olbrich}, \citenamefont
  {Zoth}, \citenamefont {Faltermeier}, \citenamefont {Lindner}, \citenamefont
  {Budkin}, \citenamefont {Tarasenko}, \citenamefont {Bel'kov}, \citenamefont
  {Kvon}, \citenamefont {Mikhailov}, \citenamefont {Dvoretsky}, \citenamefont
  {Weiss}, \citenamefont {Jenichen},\ and\ \citenamefont
  {Ganichev}}]{Dantscher2015}%
  \BibitemOpen
  \bibfield  {author} {\bibinfo {author} {\bibfnamefont {K.-M.}\ \bibnamefont
  {Dantscher}}, \bibinfo {author} {\bibfnamefont {D.~A.}\ \bibnamefont
  {Kozlov}}, \bibinfo {author} {\bibfnamefont {P.}~\bibnamefont {Olbrich}},
  \bibinfo {author} {\bibfnamefont {C.}~\bibnamefont {Zoth}}, \bibinfo {author}
  {\bibfnamefont {P.}~\bibnamefont {Faltermeier}}, \bibinfo {author}
  {\bibfnamefont {M.}~\bibnamefont {Lindner}}, \bibinfo {author} {\bibfnamefont
  {G.~V.}\ \bibnamefont {Budkin}}, \bibinfo {author} {\bibfnamefont {S.~A.}\
  \bibnamefont {Tarasenko}}, \bibinfo {author} {\bibfnamefont {V.~V.}\
  \bibnamefont {Bel'kov}}, \bibinfo {author} {\bibfnamefont {Z.~D.}\
  \bibnamefont {Kvon}}, \bibinfo {author} {\bibfnamefont {N.~N.}\ \bibnamefont
  {Mikhailov}}, \bibinfo {author} {\bibfnamefont {S.~A.}\ \bibnamefont
  {Dvoretsky}}, \bibinfo {author} {\bibfnamefont {D.}~\bibnamefont {Weiss}},
  \bibinfo {author} {\bibfnamefont {B.}~\bibnamefont {Jenichen}}, \ and\
  \bibinfo {author} {\bibfnamefont {S.~D.}\ \bibnamefont {Ganichev}},\ }\href
  {\doibase 10.1103/PhysRevB.92.165314} {\bibfield  {journal} {\bibinfo
  {journal} {Phys. Rev. B}\ }\textbf {\bibinfo {volume} {92}},\ \bibinfo
  {pages} {165314} (\bibinfo {year} {2015})}\BibitemShut {NoStop}%
\bibitem [{\citenamefont {Shuvaev}\ \emph {et~al.}(2016)\citenamefont
  {Shuvaev}, \citenamefont {Dziom}, \citenamefont {Kvon}, \citenamefont
  {Mikhailov},\ and\ \citenamefont {Pimenov}}]{Shuvaev2016}%
  \BibitemOpen
  \bibfield  {author} {\bibinfo {author} {\bibfnamefont {A.}~\bibnamefont
  {Shuvaev}}, \bibinfo {author} {\bibfnamefont {V.}~\bibnamefont {Dziom}},
  \bibinfo {author} {\bibfnamefont {Z.}~\bibnamefont {Kvon}}, \bibinfo {author}
  {\bibfnamefont {N.}~\bibnamefont {Mikhailov}}, \ and\ \bibinfo {author}
  {\bibfnamefont {A.}~\bibnamefont {Pimenov}},\ }\href {\doibase
  10.1103/physrevlett.117.117401} {\bibfield  {journal} {\bibinfo  {journal}
  {Phys. Rev. Lett.}\ }\textbf {\bibinfo {volume} {117}},\ \bibinfo {pages}
  {117401} (\bibinfo {year} {2016})}\BibitemShut {NoStop}%
\bibitem [{\citenamefont {Dantscher}\ \emph {et~al.}(2017)\citenamefont
  {Dantscher}, \citenamefont {Kozlov}, \citenamefont {Scherr}, \citenamefont
  {Gebert}, \citenamefont {Bärenfänger}, \citenamefont {Durnev},
  \citenamefont {Tarasenko}, \citenamefont {Bel'kov}, \citenamefont
  {Mikhailov}, \citenamefont {Dvoretsky}, \citenamefont {Kvon}, \citenamefont
  {Ziegler}, \citenamefont {Weiss},\ and\ \citenamefont
  {Ganichev}}]{Dantscher2017}%
  \BibitemOpen
  \bibfield  {author} {\bibinfo {author} {\bibfnamefont {K.-M.}\ \bibnamefont
  {Dantscher}}, \bibinfo {author} {\bibfnamefont {D.~A.}\ \bibnamefont
  {Kozlov}}, \bibinfo {author} {\bibfnamefont {M.~T.}\ \bibnamefont {Scherr}},
  \bibinfo {author} {\bibfnamefont {S.}~\bibnamefont {Gebert}}, \bibinfo
  {author} {\bibfnamefont {J.}~\bibnamefont {Bärenfänger}}, \bibinfo {author}
  {\bibfnamefont {M.~V.}\ \bibnamefont {Durnev}}, \bibinfo {author}
  {\bibfnamefont {S.~A.}\ \bibnamefont {Tarasenko}}, \bibinfo {author}
  {\bibfnamefont {V.~V.}\ \bibnamefont {Bel'kov}}, \bibinfo {author}
  {\bibfnamefont {N.~N.}\ \bibnamefont {Mikhailov}}, \bibinfo {author}
  {\bibfnamefont {S.~A.}\ \bibnamefont {Dvoretsky}}, \bibinfo {author}
  {\bibfnamefont {Z.~D.}\ \bibnamefont {Kvon}}, \bibinfo {author}
  {\bibfnamefont {J.}~\bibnamefont {Ziegler}}, \bibinfo {author} {\bibfnamefont
  {D.}~\bibnamefont {Weiss}}, \ and\ \bibinfo {author} {\bibfnamefont {S.~D.}\
  \bibnamefont {Ganichev}},\ }\href {\doibase 10.1103/physrevb.95.201103}
  {\bibfield  {journal} {\bibinfo  {journal} {Phys. Rev. B}\ }\textbf {\bibinfo
  {volume} {95}},\ \bibinfo {pages} {201103(R)} (\bibinfo {year}
  {2017})}\BibitemShut {NoStop}%
\bibitem [{\citenamefont {Dziom}\ \emph {et~al.}(2017)\citenamefont {Dziom},
  \citenamefont {Shuvaev}, \citenamefont {Pimenov}, \citenamefont {Astakhov},
  \citenamefont {Ames}, \citenamefont {Bendias}, \citenamefont {Böttcher},
  \citenamefont {Tkachov}, \citenamefont {Hankiewicz}, \citenamefont {Brüne},
  \citenamefont {Buhmann},\ and\ \citenamefont {Molenkamp}}]{Dziom2017}%
  \BibitemOpen
  \bibfield  {author} {\bibinfo {author} {\bibfnamefont {V.}~\bibnamefont
  {Dziom}}, \bibinfo {author} {\bibfnamefont {A.}~\bibnamefont {Shuvaev}},
  \bibinfo {author} {\bibfnamefont {A.}~\bibnamefont {Pimenov}}, \bibinfo
  {author} {\bibfnamefont {G.~V.}\ \bibnamefont {Astakhov}}, \bibinfo {author}
  {\bibfnamefont {C.}~\bibnamefont {Ames}}, \bibinfo {author} {\bibfnamefont
  {K.}~\bibnamefont {Bendias}}, \bibinfo {author} {\bibfnamefont
  {J.}~\bibnamefont {Böttcher}}, \bibinfo {author} {\bibfnamefont
  {G.}~\bibnamefont {Tkachov}}, \bibinfo {author} {\bibfnamefont {E.~M.}\
  \bibnamefont {Hankiewicz}}, \bibinfo {author} {\bibfnamefont
  {C.}~\bibnamefont {Brüne}}, \bibinfo {author} {\bibfnamefont
  {H.}~\bibnamefont {Buhmann}}, \ and\ \bibinfo {author} {\bibfnamefont
  {L.~W.}\ \bibnamefont {Molenkamp}},\ }\href {\doibase 10.1038/ncomms15197}
  {\bibfield  {journal} {\bibinfo  {journal} {Nat. Commun.}\ }\textbf {\bibinfo
  {volume} {8}},\ \bibinfo {pages} {15197} (\bibinfo {year}
  {2017})}\BibitemShut {NoStop}%
\bibitem [{\citenamefont {Kadykov}\ \emph {et~al.}(2018)\citenamefont
  {Kadykov}, \citenamefont {Krishtopenko}, \citenamefont {Jouault},
  \citenamefont {Desrat}, \citenamefont {Knap}, \citenamefont {Ruffenach},
  \citenamefont {Consejo}, \citenamefont {Torres}, \citenamefont {Morozov},
  \citenamefont {Mikhailov}, \citenamefont {Dvoretskii},\ and\ \citenamefont
  {Teppe}}]{Kadykov2018}%
  \BibitemOpen
  \bibfield  {author} {\bibinfo {author} {\bibfnamefont {A.}~\bibnamefont
  {Kadykov}}, \bibinfo {author} {\bibfnamefont {S.}~\bibnamefont
  {Krishtopenko}}, \bibinfo {author} {\bibfnamefont {B.}~\bibnamefont
  {Jouault}}, \bibinfo {author} {\bibfnamefont {W.}~\bibnamefont {Desrat}},
  \bibinfo {author} {\bibfnamefont {W.}~\bibnamefont {Knap}}, \bibinfo {author}
  {\bibfnamefont {S.}~\bibnamefont {Ruffenach}}, \bibinfo {author}
  {\bibfnamefont {C.}~\bibnamefont {Consejo}}, \bibinfo {author} {\bibfnamefont
  {J.}~\bibnamefont {Torres}}, \bibinfo {author} {\bibfnamefont
  {S.}~\bibnamefont {Morozov}}, \bibinfo {author} {\bibfnamefont
  {N.}~\bibnamefont {Mikhailov}}, \bibinfo {author} {\bibfnamefont
  {S.}~\bibnamefont {Dvoretskii}}, \ and\ \bibinfo {author} {\bibfnamefont
  {F.}~\bibnamefont {Teppe}},\ }\href {\doibase 10.1103/physrevlett.120.086401}
  {\bibfield  {journal} {\bibinfo  {journal} {Phys. Rev. Lett.}\ }\textbf
  {\bibinfo {volume} {120}},\ \bibinfo {pages} {086401} (\bibinfo {year}
  {2018})}\BibitemShut {NoStop}%
\bibitem [{\citenamefont {Gospodari{\v{c}}}\ \emph {et~al.}(2019)\citenamefont
  {Gospodari{\v{c}}}, \citenamefont {Dziom}, \citenamefont {Shuvaev},
  \citenamefont {Dobretsova}, \citenamefont {Mikhailov}, \citenamefont {Kvon},\
  and\ \citenamefont {Pimenov}}]{Gospodaric2019}%
  \BibitemOpen
  \bibfield  {author} {\bibinfo {author} {\bibfnamefont {J.}~\bibnamefont
  {Gospodari{\v{c}}}}, \bibinfo {author} {\bibfnamefont {V.}~\bibnamefont
  {Dziom}}, \bibinfo {author} {\bibfnamefont {A.}~\bibnamefont {Shuvaev}},
  \bibinfo {author} {\bibfnamefont {A.~A.}\ \bibnamefont {Dobretsova}},
  \bibinfo {author} {\bibfnamefont {N.~N.}\ \bibnamefont {Mikhailov}}, \bibinfo
  {author} {\bibfnamefont {Z.~D.}\ \bibnamefont {Kvon}}, \ and\ \bibinfo
  {author} {\bibfnamefont {A.}~\bibnamefont {Pimenov}},\ }\href {\doibase
  10.1103/physrevb.99.115130} {\bibfield  {journal} {\bibinfo  {journal} {Phys.
  Rev. B}\ }\textbf {\bibinfo {volume} {99}},\ \bibinfo {pages} {115130}
  (\bibinfo {year} {2019})}\BibitemShut {NoStop}%
\bibitem [{\citenamefont {Berchenko}\ and\ \citenamefont
  {Pashkovskii}(1976)}]{Berchenko1976}%
  \BibitemOpen
  \bibfield  {author} {\bibinfo {author} {\bibfnamefont {N.~N.}\ \bibnamefont
  {Berchenko}}\ and\ \bibinfo {author} {\bibfnamefont {M.~V.}\ \bibnamefont
  {Pashkovskii}},\ }\href@noop {} {\bibfield  {journal} {\bibinfo  {journal}
  {Sov. Phys. Usp.}\ }\textbf {\bibinfo {volume} {19}},\ \bibinfo {pages} {462}
  (\bibinfo {year} {1976})}\BibitemShut {NoStop}%
\bibitem [{\citenamefont {Orlita}\ \emph {et~al.}(2014)\citenamefont {Orlita},
  \citenamefont {Basko}, \citenamefont {Zholudev}, \citenamefont {Teppe},
  \citenamefont {Knap}, \citenamefont {Gavrilenko}, \citenamefont {Mikhailov},
  \citenamefont {Dvoretskii}, \citenamefont {Neugebauer}, \citenamefont
  {Faugeras}, \citenamefont {Barra}, \citenamefont {Martinez},\ and\
  \citenamefont {Potemski}}]{Orlita2014}%
  \BibitemOpen
  \bibfield  {author} {\bibinfo {author} {\bibfnamefont {M.}~\bibnamefont
  {Orlita}}, \bibinfo {author} {\bibfnamefont {D.~M.}\ \bibnamefont {Basko}},
  \bibinfo {author} {\bibfnamefont {M.~S.}\ \bibnamefont {Zholudev}}, \bibinfo
  {author} {\bibfnamefont {F.}~\bibnamefont {Teppe}}, \bibinfo {author}
  {\bibfnamefont {W.}~\bibnamefont {Knap}}, \bibinfo {author} {\bibfnamefont
  {V.~I.}\ \bibnamefont {Gavrilenko}}, \bibinfo {author} {\bibfnamefont
  {N.~N.}\ \bibnamefont {Mikhailov}}, \bibinfo {author} {\bibfnamefont {S.~A.}\
  \bibnamefont {Dvoretskii}}, \bibinfo {author} {\bibfnamefont
  {P.}~\bibnamefont {Neugebauer}}, \bibinfo {author} {\bibfnamefont
  {C.}~\bibnamefont {Faugeras}}, \bibinfo {author} {\bibfnamefont {A.-L.}\
  \bibnamefont {Barra}}, \bibinfo {author} {\bibfnamefont {G.}~\bibnamefont
  {Martinez}}, \ and\ \bibinfo {author} {\bibfnamefont {M.}~\bibnamefont
  {Potemski}},\ }\href {\doibase 10.1038/nphys2857} {\bibfield  {journal}
  {\bibinfo  {journal} {Nat. Phys.}\ }\textbf {\bibinfo {volume} {10}},\
  \bibinfo {pages} {233} (\bibinfo {year} {2014})}\BibitemShut {NoStop}%
\bibitem [{\citenamefont {Teppe}\ \emph {et~al.}(2016)\citenamefont {Teppe},
  \citenamefont {Marcinkiewicz}, \citenamefont {Krishtopenko}, \citenamefont
  {Ruffenach}, \citenamefont {Consejo}, \citenamefont {Kadykov}, \citenamefont
  {Desrat}, \citenamefont {But}, \citenamefont {Knap}, \citenamefont {Ludwig},
  \citenamefont {Moon}, \citenamefont {Smirnov}, \citenamefont {Orlita},
  \citenamefont {Jiang}, \citenamefont {Morozov}, \citenamefont {Gavrilenko},
  \citenamefont {Mikhailov},\ and\ \citenamefont {Dvoretskii}}]{Teppe2016}%
  \BibitemOpen
  \bibfield  {author} {\bibinfo {author} {\bibfnamefont {F.}~\bibnamefont
  {Teppe}}, \bibinfo {author} {\bibfnamefont {M.}~\bibnamefont
  {Marcinkiewicz}}, \bibinfo {author} {\bibfnamefont {S.~S.}\ \bibnamefont
  {Krishtopenko}}, \bibinfo {author} {\bibfnamefont {S.}~\bibnamefont
  {Ruffenach}}, \bibinfo {author} {\bibfnamefont {C.}~\bibnamefont {Consejo}},
  \bibinfo {author} {\bibfnamefont {A.~M.}\ \bibnamefont {Kadykov}}, \bibinfo
  {author} {\bibfnamefont {W.}~\bibnamefont {Desrat}}, \bibinfo {author}
  {\bibfnamefont {D.}~\bibnamefont {But}}, \bibinfo {author} {\bibfnamefont
  {W.}~\bibnamefont {Knap}}, \bibinfo {author} {\bibfnamefont {J.}~\bibnamefont
  {Ludwig}}, \bibinfo {author} {\bibfnamefont {S.}~\bibnamefont {Moon}},
  \bibinfo {author} {\bibfnamefont {D.}~\bibnamefont {Smirnov}}, \bibinfo
  {author} {\bibfnamefont {M.}~\bibnamefont {Orlita}}, \bibinfo {author}
  {\bibfnamefont {Z.}~\bibnamefont {Jiang}}, \bibinfo {author} {\bibfnamefont
  {S.~V.}\ \bibnamefont {Morozov}}, \bibinfo {author} {\bibfnamefont
  {V.}~\bibnamefont {Gavrilenko}}, \bibinfo {author} {\bibfnamefont {N.~N.}\
  \bibnamefont {Mikhailov}}, \ and\ \bibinfo {author} {\bibfnamefont {S.~A.}\
  \bibnamefont {Dvoretskii}},\ }\href {\doibase 10.1038/ncomms12576} {\bibfield
   {journal} {\bibinfo  {journal} {Nat. Commun.}\ }\textbf {\bibinfo {volume}
  {7}},\ \bibinfo {pages} {12576} (\bibinfo {year} {2016})}\BibitemShut
  {NoStop}%
\bibitem [{\citenamefont {Yavorskiy}\ \emph {et~al.}(2018)\citenamefont
  {Yavorskiy}, \citenamefont {Karpierz}, \citenamefont {Baj}, \citenamefont
  {B{\k{a}}k}, \citenamefont {Mikhailov}, \citenamefont {Dvoretsky},
  \citenamefont {Gavrilenko}, \citenamefont {Knap}, \citenamefont {Teppe},\
  and\ \citenamefont {Lusakowski}}]{Yavorskiy2018}%
  \BibitemOpen
  \bibfield  {author} {\bibinfo {author} {\bibfnamefont {D.}~\bibnamefont
  {Yavorskiy}}, \bibinfo {author} {\bibfnamefont {K.}~\bibnamefont {Karpierz}},
  \bibinfo {author} {\bibfnamefont {M.}~\bibnamefont {Baj}}, \bibinfo {author}
  {\bibfnamefont {M.}~\bibnamefont {B{\k{a}}k}}, \bibinfo {author}
  {\bibfnamefont {N.}~\bibnamefont {Mikhailov}}, \bibinfo {author}
  {\bibfnamefont {S.}~\bibnamefont {Dvoretsky}}, \bibinfo {author}
  {\bibfnamefont {V.}~\bibnamefont {Gavrilenko}}, \bibinfo {author}
  {\bibfnamefont {W.}~\bibnamefont {Knap}}, \bibinfo {author} {\bibfnamefont
  {F.}~\bibnamefont {Teppe}}, \ and\ \bibinfo {author} {\bibfnamefont
  {J.}~\bibnamefont {Lusakowski}},\ }\href {\doibase 10.3390/s18124341}
  {\bibfield  {journal} {\bibinfo  {journal} {Sensors}\ }\textbf {\bibinfo
  {volume} {18}},\ \bibinfo {pages} {4341} (\bibinfo {year}
  {2018})}\BibitemShut {NoStop}%
\bibitem [{\citenamefont {But}\ \emph {et~al.}(2019)\citenamefont {But},
  \citenamefont {Mittendorff}, \citenamefont {Consejo}, \citenamefont {Teppe},
  \citenamefont {Mikhailov}, \citenamefont {Dvoretskii}, \citenamefont
  {Faugeras}, \citenamefont {Winnerl}, \citenamefont {Helm}, \citenamefont
  {Knap}, \citenamefont {Potemski},\ and\ \citenamefont {Orlita}}]{But2019}%
  \BibitemOpen
  \bibfield  {author} {\bibinfo {author} {\bibfnamefont {D.~B.}\ \bibnamefont
  {But}}, \bibinfo {author} {\bibfnamefont {M.}~\bibnamefont {Mittendorff}},
  \bibinfo {author} {\bibfnamefont {C.}~\bibnamefont {Consejo}}, \bibinfo
  {author} {\bibfnamefont {F.}~\bibnamefont {Teppe}}, \bibinfo {author}
  {\bibfnamefont {N.~N.}\ \bibnamefont {Mikhailov}}, \bibinfo {author}
  {\bibfnamefont {S.~A.}\ \bibnamefont {Dvoretskii}}, \bibinfo {author}
  {\bibfnamefont {C.}~\bibnamefont {Faugeras}}, \bibinfo {author}
  {\bibfnamefont {S.}~\bibnamefont {Winnerl}}, \bibinfo {author} {\bibfnamefont
  {M.}~\bibnamefont {Helm}}, \bibinfo {author} {\bibfnamefont {W.}~\bibnamefont
  {Knap}}, \bibinfo {author} {\bibfnamefont {M.}~\bibnamefont {Potemski}}, \
  and\ \bibinfo {author} {\bibfnamefont {M.}~\bibnamefont {Orlita}},\ }\href
  {\doibase 10.1038/s41566-019-0496-1} {\bibfield  {journal} {\bibinfo
  {journal} {Nat. Photonics}\ }\textbf {\bibinfo {volume} {13}},\ \bibinfo
  {pages} {783} (\bibinfo {year} {2019})}\BibitemShut {NoStop}%
\bibitem [{\citenamefont {Tomaka}\ \emph {et~al.}(2017)\citenamefont {Tomaka},
  \citenamefont {Grendysa}, \citenamefont {Marchewka}, \citenamefont
  {{\'{S}}li{\.{z}}}, \citenamefont {Becker}, \citenamefont {Stadler},\ and\
  \citenamefont {Sheregii}}]{Tomaka2017}%
  \BibitemOpen
  \bibfield  {author} {\bibinfo {author} {\bibfnamefont {G.}~\bibnamefont
  {Tomaka}}, \bibinfo {author} {\bibfnamefont {J.}~\bibnamefont {Grendysa}},
  \bibinfo {author} {\bibfnamefont {M.}~\bibnamefont {Marchewka}}, \bibinfo
  {author} {\bibfnamefont {P.}~\bibnamefont {{\'{S}}li{\.{z}}}}, \bibinfo
  {author} {\bibfnamefont {C.}~\bibnamefont {Becker}}, \bibinfo {author}
  {\bibfnamefont {A.}~\bibnamefont {Stadler}}, \ and\ \bibinfo {author}
  {\bibfnamefont {E.}~\bibnamefont {Sheregii}},\ }\href {\doibase
  10.1016/j.opelre.2017.06.006} {\bibfield  {journal} {\bibinfo  {journal}
  {Opto-Electron. Rev.}\ }\textbf {\bibinfo {volume} {25}},\ \bibinfo {pages}
  {188} (\bibinfo {year} {2017})}\BibitemShut {NoStop}%
\bibitem [{\citenamefont {Galeeva}\ \emph {et~al.}(2018)\citenamefont
  {Galeeva}, \citenamefont {Artamkin}, \citenamefont {Kazakov}, \citenamefont
  {Danilov}, \citenamefont {Dvoretskiy}, \citenamefont {Mikhailov},
  \citenamefont {Ryabova},\ and\ \citenamefont {Khokhlov}}]{Galeeva2018}%
  \BibitemOpen
  \bibfield  {author} {\bibinfo {author} {\bibfnamefont {A.~V.}\ \bibnamefont
  {Galeeva}}, \bibinfo {author} {\bibfnamefont {A.~I.}\ \bibnamefont
  {Artamkin}}, \bibinfo {author} {\bibfnamefont {A.~S.}\ \bibnamefont
  {Kazakov}}, \bibinfo {author} {\bibfnamefont {S.~N.}\ \bibnamefont
  {Danilov}}, \bibinfo {author} {\bibfnamefont {S.~A.}\ \bibnamefont
  {Dvoretskiy}}, \bibinfo {author} {\bibfnamefont {N.~N.}\ \bibnamefont
  {Mikhailov}}, \bibinfo {author} {\bibfnamefont {L.~I.}\ \bibnamefont
  {Ryabova}}, \ and\ \bibinfo {author} {\bibfnamefont {D.~R.}\ \bibnamefont
  {Khokhlov}},\ }\href {\doibase 10.3762/bjnano.9.96} {\bibfield  {journal}
  {\bibinfo  {journal} {Beilstein J. Nanotechnol.}\ }\textbf {\bibinfo {volume}
  {9}},\ \bibinfo {pages} {1035} (\bibinfo {year} {2018})}\BibitemShut
  {NoStop}%
\bibitem [{\citenamefont {Otteneder}\ \emph {et~al.}(2018)\citenamefont
  {Otteneder}, \citenamefont {Dmitriev}, \citenamefont {Candussio},
  \citenamefont {Savchenko}, \citenamefont {Kozlov}, \citenamefont {Bel'kov},
  \citenamefont {Kvon}, \citenamefont {Mikhailov}, \citenamefont {Dvoretsky},\
  and\ \citenamefont {Ganichev}}]{Otteneder2018}%
  \BibitemOpen
  \bibfield  {author} {\bibinfo {author} {\bibfnamefont {M.}~\bibnamefont
  {Otteneder}}, \bibinfo {author} {\bibfnamefont {I.~A.}\ \bibnamefont
  {Dmitriev}}, \bibinfo {author} {\bibfnamefont {S.}~\bibnamefont {Candussio}},
  \bibinfo {author} {\bibfnamefont {M.~L.}\ \bibnamefont {Savchenko}}, \bibinfo
  {author} {\bibfnamefont {D.~A.}\ \bibnamefont {Kozlov}}, \bibinfo {author}
  {\bibfnamefont {V.~V.}\ \bibnamefont {Bel'kov}}, \bibinfo {author}
  {\bibfnamefont {Z.~D.}\ \bibnamefont {Kvon}}, \bibinfo {author}
  {\bibfnamefont {N.~N.}\ \bibnamefont {Mikhailov}}, \bibinfo {author}
  {\bibfnamefont {S.~A.}\ \bibnamefont {Dvoretsky}}, \ and\ \bibinfo {author}
  {\bibfnamefont {S.~D.}\ \bibnamefont {Ganichev}},\ }\href {\doibase
  10.1103/PhysRevB.98.245304} {\bibfield  {journal} {\bibinfo  {journal} {Phys.
  Rev. B}\ }\textbf {\bibinfo {volume} {98}},\ \bibinfo {pages} {245304}
  (\bibinfo {year} {2018})}\BibitemShut {NoStop}%
\bibitem [{\citenamefont {Candussio}\ \emph {et~al.}(2019)\citenamefont
  {Candussio}, \citenamefont {Budkin}, \citenamefont {Otteneder}, \citenamefont
  {Kozlov}, \citenamefont {Dmitriev}, \citenamefont {Bel'kov}, \citenamefont
  {Kvon}, \citenamefont {Mikhailov}, \citenamefont {Dvoretsky},\ and\
  \citenamefont {Ganichev}}]{Candussio2019}%
  \BibitemOpen
  \bibfield  {author} {\bibinfo {author} {\bibfnamefont {S.}~\bibnamefont
  {Candussio}}, \bibinfo {author} {\bibfnamefont {G.~V.}\ \bibnamefont
  {Budkin}}, \bibinfo {author} {\bibfnamefont {M.}~\bibnamefont {Otteneder}},
  \bibinfo {author} {\bibfnamefont {D.~A.}\ \bibnamefont {Kozlov}}, \bibinfo
  {author} {\bibfnamefont {I.~A.}\ \bibnamefont {Dmitriev}}, \bibinfo {author}
  {\bibfnamefont {V.~V.}\ \bibnamefont {Bel'kov}}, \bibinfo {author}
  {\bibfnamefont {Z.~D.}\ \bibnamefont {Kvon}}, \bibinfo {author}
  {\bibfnamefont {N.~N.}\ \bibnamefont {Mikhailov}}, \bibinfo {author}
  {\bibfnamefont {S.~A.}\ \bibnamefont {Dvoretsky}}, \ and\ \bibinfo {author}
  {\bibfnamefont {S.~D.}\ \bibnamefont {Ganichev}},\ }\href {\doibase
  10.1103/PhysRevMaterials.3.054205} {\bibfield  {journal} {\bibinfo  {journal}
  {Phys. Rev. Materials}\ }\textbf {\bibinfo {volume} {3}},\ \bibinfo {pages}
  {054205} (\bibinfo {year} {2019})}\BibitemShut {NoStop}%
\bibitem [{\citenamefont {Kvon}\ \emph
  {et~al.}(2008{\natexlab{a}})\citenamefont {Kvon}, \citenamefont
  {Olshanetsky}, \citenamefont {Kozlov}, \citenamefont {Mikhailov},\ and\
  \citenamefont {Dvoretskii}}]{Kvon2008}%
  \BibitemOpen
  \bibfield  {author} {\bibinfo {author} {\bibfnamefont {Z.~D.}\ \bibnamefont
  {Kvon}}, \bibinfo {author} {\bibfnamefont {E.~B.}\ \bibnamefont
  {Olshanetsky}}, \bibinfo {author} {\bibfnamefont {D.~A.}\ \bibnamefont
  {Kozlov}}, \bibinfo {author} {\bibfnamefont {N.~N.}\ \bibnamefont
  {Mikhailov}}, \ and\ \bibinfo {author} {\bibfnamefont {S.~A.}\ \bibnamefont
  {Dvoretskii}},\ }\href {\doibase 10.1134/S0021364008090117} {\bibfield
  {journal} {\bibinfo  {journal} {JETP Lett.}\ }\textbf {\bibinfo {volume}
  {87}},\ \bibinfo {pages} {502} (\bibinfo {year}
  {2008}{\natexlab{a}})}\BibitemShut {NoStop}%
\bibitem [{\citenamefont {Savchenko}\ \emph {et~al.}(2019)\citenamefont
  {Savchenko}, \citenamefont {Kozlov}, \citenamefont {Vasilev}, \citenamefont
  {Kvon}, \citenamefont {Mikhailov}, \citenamefont {Dvoretsky},\ and\
  \citenamefont {Kolesnikov}}]{Savchenko2019}%
  \BibitemOpen
  \bibfield  {author} {\bibinfo {author} {\bibfnamefont {M.~L.}\ \bibnamefont
  {Savchenko}}, \bibinfo {author} {\bibfnamefont {D.~A.}\ \bibnamefont
  {Kozlov}}, \bibinfo {author} {\bibfnamefont {N.~N.}\ \bibnamefont {Vasilev}},
  \bibinfo {author} {\bibfnamefont {Z.~D.}\ \bibnamefont {Kvon}}, \bibinfo
  {author} {\bibfnamefont {N.~N.}\ \bibnamefont {Mikhailov}}, \bibinfo {author}
  {\bibfnamefont {S.~A.}\ \bibnamefont {Dvoretsky}}, \ and\ \bibinfo {author}
  {\bibfnamefont {A.~V.}\ \bibnamefont {Kolesnikov}},\ }\href {\doibase
  10.1103/PhysRevB.99.195423} {\bibfield  {journal} {\bibinfo  {journal} {Phys.
  Rev. B}\ }\textbf {\bibinfo {volume} {99}},\ \bibinfo {pages} {195423}
  (\bibinfo {year} {2019})}\BibitemShut {NoStop}%
\bibitem [{\citenamefont {Kvon}\ \emph
  {et~al.}(2008{\natexlab{b}})\citenamefont {Kvon}, \citenamefont
  {Olshanetsky}, \citenamefont {Kozlov}, \citenamefont {Mikhailov},\ and\
  \citenamefont {Dvoretskii}}]{Kvon2008a}%
  \BibitemOpen
  \bibfield  {author} {\bibinfo {author} {\bibfnamefont {Z.~D.}\ \bibnamefont
  {Kvon}}, \bibinfo {author} {\bibfnamefont {E.~B.}\ \bibnamefont
  {Olshanetsky}}, \bibinfo {author} {\bibfnamefont {D.~A.}\ \bibnamefont
  {Kozlov}}, \bibinfo {author} {\bibfnamefont {N.~N.}\ \bibnamefont
  {Mikhailov}}, \ and\ \bibinfo {author} {\bibfnamefont {S.~A.}\ \bibnamefont
  {Dvoretskii}},\ }\href@noop {} {\bibfield  {journal} {\bibinfo  {journal}
  {JETP Lett.}\ }\textbf {\bibinfo {volume} {87}},\ \bibinfo {pages} {502}
  (\bibinfo {year} {2008}{\natexlab{b}})}\BibitemShut {NoStop}%
\bibitem [{\citenamefont {Olbrich}\ \emph {et~al.}(2011)\citenamefont
  {Olbrich}, \citenamefont {Karch}, \citenamefont {Ivchenko}, \citenamefont
  {Kamann}, \citenamefont {März}, \citenamefont {Fehrenbacher}, \citenamefont
  {Weiss},\ and\ \citenamefont {Ganichev}}]{Olbrich2011}%
  \BibitemOpen
  \bibfield  {author} {\bibinfo {author} {\bibfnamefont {P.}~\bibnamefont
  {Olbrich}}, \bibinfo {author} {\bibfnamefont {J.}~\bibnamefont {Karch}},
  \bibinfo {author} {\bibfnamefont {E.~L.}\ \bibnamefont {Ivchenko}}, \bibinfo
  {author} {\bibfnamefont {J.}~\bibnamefont {Kamann}}, \bibinfo {author}
  {\bibfnamefont {B.}~\bibnamefont {März}}, \bibinfo {author} {\bibfnamefont
  {M.}~\bibnamefont {Fehrenbacher}}, \bibinfo {author} {\bibfnamefont
  {D.}~\bibnamefont {Weiss}}, \ and\ \bibinfo {author} {\bibfnamefont {S.~D.}\
  \bibnamefont {Ganichev}},\ }\href {\doibase 10.1103/physrevb.83.165320}
  {\bibfield  {journal} {\bibinfo  {journal} {Phys. Rev. B}\ }\textbf {\bibinfo
  {volume} {83}},\ \bibinfo {pages} {165320} (\bibinfo {year}
  {2011})}\BibitemShut {NoStop}%
\bibitem [{\citenamefont {Ganichev}\ \emph
  {et~al.}(2002{\natexlab{a}})\citenamefont {Ganichev}, \citenamefont
  {Rössler}, \citenamefont {Prettl}, \citenamefont {Ivchenko}, \citenamefont
  {Bel'kov}, \citenamefont {Neumann}, \citenamefont {Brunner},\ and\
  \citenamefont {Abstreiter}}]{Ganichev2002a}%
  \BibitemOpen
  \bibfield  {author} {\bibinfo {author} {\bibfnamefont {S.~D.}\ \bibnamefont
  {Ganichev}}, \bibinfo {author} {\bibfnamefont {U.}~\bibnamefont {Rössler}},
  \bibinfo {author} {\bibfnamefont {W.}~\bibnamefont {Prettl}}, \bibinfo
  {author} {\bibfnamefont {E.~L.}\ \bibnamefont {Ivchenko}}, \bibinfo {author}
  {\bibfnamefont {V.~V.}\ \bibnamefont {Bel'kov}}, \bibinfo {author}
  {\bibfnamefont {R.}~\bibnamefont {Neumann}}, \bibinfo {author} {\bibfnamefont
  {K.}~\bibnamefont {Brunner}}, \ and\ \bibinfo {author} {\bibfnamefont
  {G.}~\bibnamefont {Abstreiter}},\ }\href {\doibase
  10.1103/physrevb.66.075328} {\bibfield  {journal} {\bibinfo  {journal} {Phys.
  Rev. B}\ }\textbf {\bibinfo {volume} {66}},\ \bibinfo {pages} {075328}
  (\bibinfo {year} {2002}{\natexlab{a}})}\BibitemShut {NoStop}%
\bibitem [{\citenamefont {Ganichev}\ \emph
  {et~al.}(2002{\natexlab{b}})\citenamefont {Ganichev}, \citenamefont
  {Danilov}, \citenamefont {Bel'kov}, \citenamefont {Ivchenko}, \citenamefont
  {Bichler}, \citenamefont {Wegscheider}, \citenamefont {Weiss},\ and\
  \citenamefont {Prettl}}]{Ganichev2002b}%
  \BibitemOpen
  \bibfield  {author} {\bibinfo {author} {\bibfnamefont {S.~D.}\ \bibnamefont
  {Ganichev}}, \bibinfo {author} {\bibfnamefont {S.~N.}\ \bibnamefont
  {Danilov}}, \bibinfo {author} {\bibfnamefont {V.~V.}\ \bibnamefont
  {Bel'kov}}, \bibinfo {author} {\bibfnamefont {E.~L.}\ \bibnamefont
  {Ivchenko}}, \bibinfo {author} {\bibfnamefont {M.}~\bibnamefont {Bichler}},
  \bibinfo {author} {\bibfnamefont {W.}~\bibnamefont {Wegscheider}}, \bibinfo
  {author} {\bibfnamefont {D.}~\bibnamefont {Weiss}}, \ and\ \bibinfo {author}
  {\bibfnamefont {W.}~\bibnamefont {Prettl}},\ }\href {\doibase
  10.1103/physrevlett.88.057401} {\bibfield  {journal} {\bibinfo  {journal}
  {Phys. Rev. Lett.}\ }\textbf {\bibinfo {volume} {88}},\ \bibinfo {pages}
  {057401} (\bibinfo {year} {2002}{\natexlab{b}})}\BibitemShut {NoStop}%
\bibitem [{\citenamefont {Weber}\ \emph {et~al.}(2008)\citenamefont {Weber},
  \citenamefont {Golub}, \citenamefont {Danilov}, \citenamefont {Karch},
  \citenamefont {Reitmaier}, \citenamefont {Wittmann}, \citenamefont {Bel'kov},
  \citenamefont {Ivchenko}, \citenamefont {Kvon}, \citenamefont {Vinh},
  \citenamefont {van~der Meer}, \citenamefont {Murdin},\ and\ \citenamefont
  {Ganichev}}]{Weber2008}%
  \BibitemOpen
  \bibfield  {author} {\bibinfo {author} {\bibfnamefont {W.}~\bibnamefont
  {Weber}}, \bibinfo {author} {\bibfnamefont {L.~E.}\ \bibnamefont {Golub}},
  \bibinfo {author} {\bibfnamefont {S.~N.}\ \bibnamefont {Danilov}}, \bibinfo
  {author} {\bibfnamefont {J.}~\bibnamefont {Karch}}, \bibinfo {author}
  {\bibfnamefont {C.}~\bibnamefont {Reitmaier}}, \bibinfo {author}
  {\bibfnamefont {B.}~\bibnamefont {Wittmann}}, \bibinfo {author}
  {\bibfnamefont {V.~V.}\ \bibnamefont {Bel'kov}}, \bibinfo {author}
  {\bibfnamefont {E.~L.}\ \bibnamefont {Ivchenko}}, \bibinfo {author}
  {\bibfnamefont {Z.~D.}\ \bibnamefont {Kvon}}, \bibinfo {author}
  {\bibfnamefont {N.~Q.}\ \bibnamefont {Vinh}}, \bibinfo {author}
  {\bibfnamefont {A.~F.~G.}\ \bibnamefont {van~der Meer}}, \bibinfo {author}
  {\bibfnamefont {B.}~\bibnamefont {Murdin}}, \ and\ \bibinfo {author}
  {\bibfnamefont {S.~D.}\ \bibnamefont {Ganichev}},\ }\href {\doibase
  10.1103/PhysRevB.77.245304} {\bibfield  {journal} {\bibinfo  {journal} {Phys.
  Rev. B}\ }\textbf {\bibinfo {volume} {77}},\ \bibinfo {pages} {245304}
  (\bibinfo {year} {2008})}\BibitemShut {NoStop}%
\bibitem [{\citenamefont {Drexler}\ \emph {et~al.}(2012)\citenamefont
  {Drexler}, \citenamefont {Dyakonova}, \citenamefont {Olbrich}, \citenamefont
  {Karch}, \citenamefont {Schafberger}, \citenamefont {Karpierz}, \citenamefont
  {Mityagin}, \citenamefont {Lifshits}, \citenamefont {Teppe}, \citenamefont
  {Klimenko}, \citenamefont {Meziani}, \citenamefont {Knap},\ and\
  \citenamefont {Ganichev}}]{Drexler2012}%
  \BibitemOpen
  \bibfield  {author} {\bibinfo {author} {\bibfnamefont {C.}~\bibnamefont
  {Drexler}}, \bibinfo {author} {\bibfnamefont {N.}~\bibnamefont {Dyakonova}},
  \bibinfo {author} {\bibfnamefont {P.}~\bibnamefont {Olbrich}}, \bibinfo
  {author} {\bibfnamefont {J.}~\bibnamefont {Karch}}, \bibinfo {author}
  {\bibfnamefont {M.}~\bibnamefont {Schafberger}}, \bibinfo {author}
  {\bibfnamefont {K.}~\bibnamefont {Karpierz}}, \bibinfo {author}
  {\bibfnamefont {Y.}~\bibnamefont {Mityagin}}, \bibinfo {author}
  {\bibfnamefont {M.~B.}\ \bibnamefont {Lifshits}}, \bibinfo {author}
  {\bibfnamefont {F.}~\bibnamefont {Teppe}}, \bibinfo {author} {\bibfnamefont
  {O.}~\bibnamefont {Klimenko}}, \bibinfo {author} {\bibfnamefont {Y.~M.}\
  \bibnamefont {Meziani}}, \bibinfo {author} {\bibfnamefont {W.}~\bibnamefont
  {Knap}}, \ and\ \bibinfo {author} {\bibfnamefont {S.~D.}\ \bibnamefont
  {Ganichev}},\ }\href {\doibase 10.1063/1.4729043} {\bibfield  {journal}
  {\bibinfo  {journal} {J. Appl. Phys.}\ }\textbf {\bibinfo {volume} {111}},\
  \bibinfo {pages} {124504} (\bibinfo {year} {2012})}\BibitemShut {NoStop}%
\bibitem [{\citenamefont {Kozlov}\ \emph {et~al.}(2011)\citenamefont {Kozlov},
  \citenamefont {Kvon}, \citenamefont {Mikhailov}, \citenamefont {Dvoretskii},\
  and\ \citenamefont {Portal}}]{Kozlov2011}%
  \BibitemOpen
  \bibfield  {author} {\bibinfo {author} {\bibfnamefont {D.~A.}\ \bibnamefont
  {Kozlov}}, \bibinfo {author} {\bibfnamefont {Z.~D.}\ \bibnamefont {Kvon}},
  \bibinfo {author} {\bibfnamefont {N.~N.}\ \bibnamefont {Mikhailov}}, \bibinfo
  {author} {\bibfnamefont {S.~A.}\ \bibnamefont {Dvoretskii}}, \ and\ \bibinfo
  {author} {\bibfnamefont {J.~C.}\ \bibnamefont {Portal}},\ }\href {\doibase
  10.1134/s0021364011030088} {\bibfield  {journal} {\bibinfo  {journal} {{JETP}
  Letters}\ }\textbf {\bibinfo {volume} {93}},\ \bibinfo {pages} {170}
  (\bibinfo {year} {2011})}\BibitemShut {NoStop}%
\bibitem [{\citenamefont {Abstreiter}\ \emph {et~al.}(1976)\citenamefont
  {Abstreiter}, \citenamefont {Kotthaus}, \citenamefont {Koch},\ and\
  \citenamefont {Dorda}}]{Abstreiter1976}%
  \BibitemOpen
  \bibfield  {author} {\bibinfo {author} {\bibfnamefont {G.}~\bibnamefont
  {Abstreiter}}, \bibinfo {author} {\bibfnamefont {J.~P.}\ \bibnamefont
  {Kotthaus}}, \bibinfo {author} {\bibfnamefont {J.~F.}\ \bibnamefont {Koch}},
  \ and\ \bibinfo {author} {\bibfnamefont {G.}~\bibnamefont {Dorda}},\ }\href
  {\doibase 10.1103/physrevb.14.2480} {\bibfield  {journal} {\bibinfo
  {journal} {Phys. Rev. B}\ }\textbf {\bibinfo {volume} {14}},\ \bibinfo
  {pages} {2480} (\bibinfo {year} {1976})}\BibitemShut {NoStop}%
\bibitem [{\citenamefont {Herrmann}\ \emph {et~al.}(2016)\citenamefont
  {Herrmann}, \citenamefont {Dmitriev}, \citenamefont {Kozlov}, \citenamefont
  {Schneider}, \citenamefont {Jentzsch}, \citenamefont {Kvon}, \citenamefont
  {Olbrich}, \citenamefont {Bel'kov}, \citenamefont {Bayer}, \citenamefont
  {Schuh}, \citenamefont {Bougeard}, \citenamefont {Kuczmik}, \citenamefont
  {Oltscher}, \citenamefont {Weiss},\ and\ \citenamefont
  {Ganichev}}]{Hermann2016}%
  \BibitemOpen
  \bibfield  {author} {\bibinfo {author} {\bibfnamefont {T.}~\bibnamefont
  {Herrmann}}, \bibinfo {author} {\bibfnamefont {I.~A.}\ \bibnamefont
  {Dmitriev}}, \bibinfo {author} {\bibfnamefont {D.~A.}\ \bibnamefont
  {Kozlov}}, \bibinfo {author} {\bibfnamefont {M.}~\bibnamefont {Schneider}},
  \bibinfo {author} {\bibfnamefont {B.}~\bibnamefont {Jentzsch}}, \bibinfo
  {author} {\bibfnamefont {Z.~D.}\ \bibnamefont {Kvon}}, \bibinfo {author}
  {\bibfnamefont {P.}~\bibnamefont {Olbrich}}, \bibinfo {author} {\bibfnamefont
  {V.~V.}\ \bibnamefont {Bel'kov}}, \bibinfo {author} {\bibfnamefont
  {A.}~\bibnamefont {Bayer}}, \bibinfo {author} {\bibfnamefont
  {D.}~\bibnamefont {Schuh}}, \bibinfo {author} {\bibfnamefont
  {D.}~\bibnamefont {Bougeard}}, \bibinfo {author} {\bibfnamefont
  {T.}~\bibnamefont {Kuczmik}}, \bibinfo {author} {\bibfnamefont
  {M.}~\bibnamefont {Oltscher}}, \bibinfo {author} {\bibfnamefont
  {D.}~\bibnamefont {Weiss}}, \ and\ \bibinfo {author} {\bibfnamefont {S.~D.}\
  \bibnamefont {Ganichev}},\ }\href {\doibase 10.1103/PhysRevB.94.081301}
  {\bibfield  {journal} {\bibinfo  {journal} {Phys. Rev. B}\ }\textbf {\bibinfo
  {volume} {94}},\ \bibinfo {pages} {081301} (\bibinfo {year}
  {2016})}\BibitemShut {NoStop}%
\bibitem [{\citenamefont {Rigaux}(1980)}]{Rigaux1980}%
  \BibitemOpen
  \bibfield  {author} {\bibinfo {author} {\bibfnamefont {C.}~\bibnamefont
  {Rigaux}},\ }in\ \href {\doibase 10.1007/3-540-10261-2_38} {\emph {\bibinfo
  {booktitle} {Narrow Gap Semiconductors Physics and Applications}}}\ (\bibinfo
   {publisher} {Springer Berlin Heidelberg},\ \bibinfo {year} {1980})\ pp.\
  \bibinfo {pages} {110--124}\BibitemShut {NoStop}%
\bibitem [{\citenamefont {Dyakonov}\ \emph {et~al.}(1969)\citenamefont
  {Dyakonov}, \citenamefont {Efros},\ and\ \citenamefont
  {Mitchell}}]{Dyakonov1969}%
  \BibitemOpen
  \bibfield  {author} {\bibinfo {author} {\bibfnamefont {M.~I.}\ \bibnamefont
  {Dyakonov}}, \bibinfo {author} {\bibfnamefont {A.~L.}\ \bibnamefont {Efros}},
  \ and\ \bibinfo {author} {\bibfnamefont {D.~L.}\ \bibnamefont {Mitchell}},\
  }\href {\doibase 10.1103/physrev.180.813} {\bibfield  {journal} {\bibinfo
  {journal} {Phys. Rev.}\ }\textbf {\bibinfo {volume} {180}},\ \bibinfo {pages}
  {813} (\bibinfo {year} {1969})}\BibitemShut {NoStop}%
\bibitem [{\citenamefont {Goldman}\ \emph {et~al.}(1986)\citenamefont
  {Goldman}, \citenamefont {Drew}, \citenamefont {Shayegan},\ and\
  \citenamefont {Nelson}}]{Goldman1986}%
  \BibitemOpen
  \bibfield  {author} {\bibinfo {author} {\bibfnamefont {V.~J.}\ \bibnamefont
  {Goldman}}, \bibinfo {author} {\bibfnamefont {H.~D.}\ \bibnamefont {Drew}},
  \bibinfo {author} {\bibfnamefont {M.}~\bibnamefont {Shayegan}}, \ and\
  \bibinfo {author} {\bibfnamefont {D.~A.}\ \bibnamefont {Nelson}},\ }\href
  {\doibase 10.1103/physrevlett.56.968} {\bibfield  {journal} {\bibinfo
  {journal} {Phys. Rev. Lett.}\ }\textbf {\bibinfo {volume} {56}},\ \bibinfo
  {pages} {968} (\bibinfo {year} {1986})}\BibitemShut {NoStop}%
\bibitem [{\citenamefont {Lifshitz}\ and\ \citenamefont
  {Nad'}(1965)}]{Lifshitz1965}%
  \BibitemOpen
  \bibfield  {author} {\bibinfo {author} {\bibfnamefont {T.~M.}\ \bibnamefont
  {Lifshitz}}\ and\ \bibinfo {author} {\bibfnamefont {Y.~F.}\ \bibnamefont
  {Nad'}},\ }\href
  {http://www.mathnet.ru/php/getFT.phtml?jrnid=dan&paperid=31186&what=fullt&option_lang=eng}
  {\bibfield  {journal} {\bibinfo  {journal} {Dokl. Akad. Nauk SSSR}\ }\textbf
  {\bibinfo {volume} {162}},\ \bibinfo {pages} {801} (\bibinfo {year}
  {1965})},\ \bibinfo {note} {[Sov. Phys.-Doklady \textbf{10}, 532
  (1965)]}\BibitemShut {NoStop}%
\bibitem [{\citenamefont {Gershenzon}\ \emph {et~al.}(1973)\citenamefont
  {Gershenzon}, \citenamefont {Gol'tsman},\ and\ \citenamefont
  {Ptitsina}}]{Gershenzon1973}%
  \BibitemOpen
  \bibfield  {author} {\bibinfo {author} {\bibfnamefont {E.~M.}\ \bibnamefont
  {Gershenzon}}, \bibinfo {author} {\bibfnamefont {G.~N.}\ \bibnamefont
  {Gol'tsman}}, \ and\ \bibinfo {author} {\bibfnamefont {N.~G.}\ \bibnamefont
  {Ptitsina}},\ }\href {http://jetp.ac.ru/cgi-bin/dn/e_037_02_0299.pdf}
  {\bibfield  {journal} {\bibinfo  {journal} {Zh. Eksp. Teor. Fiz.}\ }\textbf
  {\bibinfo {volume} {64}},\ \bibinfo {pages} {587} (\bibinfo {year} {1973})},\
  \bibinfo {note} {[Sov. Phys. JETP \textbf{37}, 299--304 (1973)]}\BibitemShut
  {NoStop}%
\bibitem [{\citenamefont {Volkov}\ and\ \citenamefont
  {Pankratov}(1985)}]{Volkov1985}%
  \BibitemOpen
  \bibfield  {author} {\bibinfo {author} {\bibfnamefont {B.}~\bibnamefont
  {Volkov}}\ and\ \bibinfo {author} {\bibfnamefont {O.}~\bibnamefont
  {Pankratov}},\ }\href@noop {} {\bibfield  {journal} {\bibinfo  {journal}
  {Pis'ma Zh. Eksp. Teor. Fiz.}\ }\textbf {\bibinfo {volume} {42}},\ \bibinfo
  {pages} {145} (\bibinfo {year} {1985})},\ \bibinfo {note} {[JETP Lett.
  \textbf{42}, 178 (1985)]}\BibitemShut {NoStop}%
\bibitem [{\citenamefont {Tchoumakov}\ \emph {et~al.}(2017)\citenamefont
  {Tchoumakov}, \citenamefont {Jouffrey}, \citenamefont {Inhofer},
  \citenamefont {Bocquillon}, \citenamefont {Pla\ifmmode~\mbox{\c{c}}\else
  \c{c}\fi{}ais}, \citenamefont {Carpentier},\ and\ \citenamefont
  {Goerbig}}]{Tchoumakov2017}%
  \BibitemOpen
  \bibfield  {author} {\bibinfo {author} {\bibfnamefont {S.}~\bibnamefont
  {Tchoumakov}}, \bibinfo {author} {\bibfnamefont {V.}~\bibnamefont
  {Jouffrey}}, \bibinfo {author} {\bibfnamefont {A.}~\bibnamefont {Inhofer}},
  \bibinfo {author} {\bibfnamefont {E.}~\bibnamefont {Bocquillon}}, \bibinfo
  {author} {\bibfnamefont {B.}~\bibnamefont {Pla\ifmmode~\mbox{\c{c}}\else
  \c{c}\fi{}ais}}, \bibinfo {author} {\bibfnamefont {D.}~\bibnamefont
  {Carpentier}}, \ and\ \bibinfo {author} {\bibfnamefont {M.~O.}\ \bibnamefont
  {Goerbig}},\ }\href {\doibase 10.1103/PhysRevB.96.201302} {\bibfield
  {journal} {\bibinfo  {journal} {Phys. Rev. B}\ }\textbf {\bibinfo {volume}
  {96}},\ \bibinfo {pages} {201302} (\bibinfo {year} {2017})}\BibitemShut
  {NoStop}%
\bibitem [{\citenamefont {Hancock}\ \emph {et~al.}(2011)\citenamefont
  {Hancock}, \citenamefont {van Mechelen}, \citenamefont {Kuzmenko},
  \citenamefont {van~der Marel}, \citenamefont {Brüne}, \citenamefont {Novik},
  \citenamefont {Astakhov}, \citenamefont {Buhmann},\ and\ \citenamefont
  {Molenkamp}}]{Hancock2011}%
  \BibitemOpen
  \bibfield  {author} {\bibinfo {author} {\bibfnamefont {J.~N.}\ \bibnamefont
  {Hancock}}, \bibinfo {author} {\bibfnamefont {J.~L.~M.}\ \bibnamefont {van
  Mechelen}}, \bibinfo {author} {\bibfnamefont {A.~B.}\ \bibnamefont
  {Kuzmenko}}, \bibinfo {author} {\bibfnamefont {D.}~\bibnamefont {van~der
  Marel}}, \bibinfo {author} {\bibfnamefont {C.}~\bibnamefont {Brüne}},
  \bibinfo {author} {\bibfnamefont {E.~G.}\ \bibnamefont {Novik}}, \bibinfo
  {author} {\bibfnamefont {G.~V.}\ \bibnamefont {Astakhov}}, \bibinfo {author}
  {\bibfnamefont {H.}~\bibnamefont {Buhmann}}, \ and\ \bibinfo {author}
  {\bibfnamefont {L.~W.}\ \bibnamefont {Molenkamp}},\ }\href {\doibase
  10.1103/physrevlett.107.136803} {\bibfield  {journal} {\bibinfo  {journal}
  {Phys. Rev. Lett.}\ }\textbf {\bibinfo {volume} {107}},\ \bibinfo {pages}
  {136803} (\bibinfo {year} {2011})}\BibitemShut {NoStop}%
\bibitem [{\citenamefont {Kibis}\ \emph {et~al.}(2019)\citenamefont {Kibis},
  \citenamefont {Kyriienko},\ and\ \citenamefont {Shelykh}}]{Kibis2019}%
  \BibitemOpen
  \bibfield  {author} {\bibinfo {author} {\bibfnamefont {O.~V.}\ \bibnamefont
  {Kibis}}, \bibinfo {author} {\bibfnamefont {O.}~\bibnamefont {Kyriienko}}, \
  and\ \bibinfo {author} {\bibfnamefont {I.~A.}\ \bibnamefont {Shelykh}},\
  }\href {\doibase 10.1088/1367-2630/ab1406} {\bibfield  {journal} {\bibinfo
  {journal} {New Journal of Physics}\ }\textbf {\bibinfo {volume} {21}},\
  \bibinfo {pages} {043016} (\bibinfo {year} {2019})}\BibitemShut {NoStop}%
\bibitem [{\citenamefont {Pankratov}\ \emph {et~al.}(1987)\citenamefont
  {Pankratov}, \citenamefont {Pakhomov},\ and\ \citenamefont
  {Volkov}}]{Pankratov1987}%
  \BibitemOpen
  \bibfield  {author} {\bibinfo {author} {\bibfnamefont {O.~A.}\ \bibnamefont
  {Pankratov}}, \bibinfo {author} {\bibfnamefont {S.~V.}\ \bibnamefont
  {Pakhomov}}, \ and\ \bibinfo {author} {\bibfnamefont {B.~A.}\ \bibnamefont
  {Volkov}},\ }\href {\doibase 10.1016/0038-1098(87)90934-3} {\bibfield
  {journal} {\bibinfo  {journal} {Solid State Communications}\ }\textbf
  {\bibinfo {volume} {61}},\ \bibinfo {pages} {93} (\bibinfo {year}
  {1987})}\BibitemShut {NoStop}%
\bibitem [{\citenamefont {Bir}\ and\ \citenamefont {Pikus}(1974)}]{Bir1974}%
  \BibitemOpen
  \bibfield  {author} {\bibinfo {author} {\bibfnamefont {G.}~\bibnamefont
  {Bir}}\ and\ \bibinfo {author} {\bibfnamefont {G.}~\bibnamefont {Pikus}},\
  }\href@noop {} {\emph {\bibinfo {title} {Symmetry and Strain-induced Effects
  in Semiconductors}}}\ (\bibinfo  {publisher} {Wiley},\ \bibinfo {address}
  {New York},\ \bibinfo {year} {1974})\BibitemShut {NoStop}%
\bibitem [{\citenamefont {Novik}\ \emph {et~al.}(2005)\citenamefont {Novik},
  \citenamefont {Pfeuffer-Jeschke}, \citenamefont {Jungwirth}, \citenamefont
  {Latussek}, \citenamefont {Becker}, \citenamefont {Landwehr}, \citenamefont
  {Buhmann},\ and\ \citenamefont {Molenkamp}}]{Novik2005}%
  \BibitemOpen
  \bibfield  {author} {\bibinfo {author} {\bibfnamefont {E.~G.}\ \bibnamefont
  {Novik}}, \bibinfo {author} {\bibfnamefont {A.}~\bibnamefont
  {Pfeuffer-Jeschke}}, \bibinfo {author} {\bibfnamefont {T.}~\bibnamefont
  {Jungwirth}}, \bibinfo {author} {\bibfnamefont {V.}~\bibnamefont {Latussek}},
  \bibinfo {author} {\bibfnamefont {C.~R.}\ \bibnamefont {Becker}}, \bibinfo
  {author} {\bibfnamefont {G.}~\bibnamefont {Landwehr}}, \bibinfo {author}
  {\bibfnamefont {H.}~\bibnamefont {Buhmann}}, \ and\ \bibinfo {author}
  {\bibfnamefont {L.~W.}\ \bibnamefont {Molenkamp}},\ }\href {\doibase
  10.1103/PhysRevB.72.035321} {\bibfield  {journal} {\bibinfo  {journal} {Phys.
  Rev. B}\ }\textbf {\bibinfo {volume} {72}},\ \bibinfo {pages} {035321}
  (\bibinfo {year} {2005})}\BibitemShut {NoStop}%
\bibitem [{\citenamefont {Laurenti}\ \emph {et~al.}(1990)\citenamefont
  {Laurenti}, \citenamefont {Camassel}, \citenamefont {Bouhemadou},
  \citenamefont {Toulouse}, \citenamefont {Legros},\ and\ \citenamefont
  {Lusson}}]{Laurenti1990}%
  \BibitemOpen
  \bibfield  {author} {\bibinfo {author} {\bibfnamefont {J.~P.}\ \bibnamefont
  {Laurenti}}, \bibinfo {author} {\bibfnamefont {J.}~\bibnamefont {Camassel}},
  \bibinfo {author} {\bibfnamefont {A.}~\bibnamefont {Bouhemadou}}, \bibinfo
  {author} {\bibfnamefont {B.}~\bibnamefont {Toulouse}}, \bibinfo {author}
  {\bibfnamefont {R.}~\bibnamefont {Legros}}, \ and\ \bibinfo {author}
  {\bibfnamefont {A.}~\bibnamefont {Lusson}},\ }\href {\doibase
  10.1063/1.345119} {\bibfield  {journal} {\bibinfo  {journal} {J. Appl.
  Phys.}\ }\textbf {\bibinfo {volume} {67}},\ \bibinfo {pages} {6454} (\bibinfo
  {year} {1990})}\BibitemShut {NoStop}%
\end{thebibliography}%

\end{document}